\documentclass[12pt]{iopart}

\usepackage{setstack,amssymb,amsfonts,graphicx,iopams}	
\newcommand{\bmath}[1]{\mbox{\boldmath$#1$}}
\begin{document}
\title[Derivation of Green's Function of Spin Calogero-Sutherland Model]{%
Derivation of Green's Function of Spin Calogero-Sutherland Model\\
by Uglov's Method
}

\author{%
Ryota Nakai$^1$ and 
Yusuke Kato$^2$
}
\address{
$^1$ Department of Physics, University of Tokyo, Tokyo, Japan 113-0033}
\address{
$^2$ Department of Basic Science, University of Tokyo, Tokyo, Japan 153-8902}
\ead{rnakai@vortex.c.u-tokyo.ac.jp}
\ead{yusuke@phys.c.u-tokyo.ac.jp}
%
%
\begin{abstract}
Hole propagator of spin $1/2$ Calogero-Sutherland model
is derived using Uglov's method,
which maps the exact eigenfunctions of the model,
called Yangian Gelfand-Zetlin basis,
to a limit of Macdonald polynomials (gl$_2$-Jack polynomials).
To apply this mapping method to the calculation of 1-particle Green's function,
we confirm that the sum of the field annihilation operator $\hat{\psi}_\uparrow +\hat{\psi}_\downarrow$ on 
Yangian Gelfand-Zetlin basis 
is transformed to the field annihilation operator $\hat{\psi}$ on gl$_2$-Jack polynomials by the mapping. The resultant expression for hole propagator for finite-size system is written in terms of renormalized momenta and spin of quasi-holes and the expression in the thermodynamic limit coincides with the earlier result derived by another method. We also discuss the singularity of the spectral function for a specific coupling parameter where the hole propagator of spin Calogero-Sutherland model becomes equivalent to dynamical colour correlation function of SU(3) Haldane-Shastry model.
\end{abstract}
\submitto{\JPA}
\maketitle

\section{Introduction}

Calogero-Sutherland model\cite{Ca1,Ca2,Su1,Su2,Su3,Su4,olshanetsky83,Subook,Shi}  is 
a one-dimensional quantum system of particles with two-body interaction inversely proportional to the square of distance. 
From theoretical point of view,
Calogero-Sutherland model has attracted extensive attention
in relation to fractional exclusion statistics\cite{haldane91a,wu94,wu95}, Jack polynomials\cite{Sta,Mac}, Tomonaga-Luttinger liquid\cite{KawakamiYang91}, collective field theory\cite{Awata,Abanov} and matrix models\cite{Sim}. 
An intriguing property of Calogero-Sutherland model lies in the fact that 
exact expressions of two-point dynamical correlation functions have been obtained for whole range of time and space\cite{Sim,Min94,For,Hal2,Ha94,Ha95a,Les,Ha95b,zirnbauer95,Ser}. The expressions for dynamical correlation functions are much simpler than those of Bethe solvable models\cite{korepin93}. Through exact explicit expressions of spectral functions of Calogero-Sutherland model, dynamical properties of one-dimensional quantum systems have been discussed\cite{mucciolo,pustilnik}.

In this paper, we consider spin generalization 
of Calogero-Sutherland model\cite{Ha2,Kaw,Min},
which is the model of particles with spin 1/2 as internal degrees of freedom. 
In the following we call this model ^^ ^^ spin Calogero-Sutherland model" while the original model ^^ ^^ scalar Calogero-Sutherland model". 
Spin Calogero-Sutherland models are even more important than scalar model in the sense that (i) the spin Calogero-Sutherland models have Yangian symmetry\cite{Dri,Ber} as an internal symmetry, (ii) it reduces to the Haldane-Shastry model\cite{HS1,HS2} in the limit of infinite coupling parameter\cite{polychronakos93,sutherlandshastry93} and (iii) it realizes the Tomonaga-Luttinger liquid (of particles with spin internal symmetry) in the systems with finite number of particles in the simplest manner\cite{Kaw,kawakami93a}.  

Exact results of dynamical correlation functions such as hole propagator\cite{Kat,Ka2}, density correlation\cite{Ugl,Yam} and spin correlation function\cite{Ugl,Yam,Yamamoto2000,Yamamoto1999} have been obtained in spin Calogero-Sutherland model. However particle propagator of this model has not been derived.
With knowledge of hole and particle propagators, we can arrive at full understanding of the dynamics and elementary excitations. From the hole and particle propagators, furthermore, we can construct 1-particle causal Green's function, which provides a nontrivial starting point of many-body perturbation theory.   
The derivation of exact results applicable both to hole and particle propagators, therefore, is highly required from the above point of view. 
We address this issue in the present paper. 

There are two types of the expression of the wave functions
of the spin Calogero-Sutherland model.
The first one is given by
Jack polynomial with prescribed symmetry\cite{Bak,Dun}.
This polynomial is constructed by partially symmetrization
or anti-symmetrization of 
non-symmetric Jack polynomials\cite{opdam95,sahi96}, which are the simultaneous
eigenfunctions of the integrals of motion of this model, 
that is Cherednik-Dunkl operators\cite{Dun2,Che}.
Jack polynomials with prescribed symmetry
form an orthogonal basis of the Hilbert space with 
a specific spin configuration. 
The second one is related to the 
Yangian symmetry\cite{Dri,Ber} of spin Calogero-Sutherland model.
An orthogonal basis of 
the Fock space including spin degrees of freedom with fixed particle number is called Yangian Gelfand-Zetlin basis\cite{Naz,Tak}  

Dynamical correlation functions of spin Calogero-Sutherland model were
calculated in two methods according to the two types of eigenfunctions.
One way uses
Jack polynomials with prescribed symmetry as eigenfunctions,
and the dynamical correlation functions are
calculated by using 
 the relations derived from the non-symmetric Jack polynomials.
With this method, hole propagator
\cite{Kat,Ka2} has been derived by one of the authors and his collaborator. Recently\cite{KuraKato}, it was found that dynamical density correlation of spin Calogero-Sutherland model can be derived with Jack polynomials with prescribed symmetry and the method used in \cite{arikawasaiga}. However, no formulas necessary to particle propagator have been obtained in the theory of Jack polynomials with prescribed symmetry. 

The other way uses Yangian Gelfand-Zetlin basis 
as eigenfunctions
together with the mapping from Yangian Gelfand-Zetlin basis
to symmetric polynomials\cite{Ugl}. 
The relations of this polynomial necessary to calculate
dynamical correlation function are derived from 
those of Macdonald polynomials\cite{Mac}.
Density correlation function and spin correlation function 
have been derived with this method\cite{Ugl,Yam}.
We refer to the latter method as Uglov's method. 
For a decade, it has been unresolved issue whether the Uglov's method is applicable to the calculation of 1-particle Green's function. 
In Uglov's method, wavefunctions of multi-component particles are mapped to those of single-component particles. 
It is not obvious how the field operator in 1-particle Green's function
is transformed under this mapping, in contrast to the density operator or spin operator considered in \cite{Ugl}. 

The main purpose of this paper is to show 
that Uglov's method is applicable to derive both hole and particle propagators. More explicitly, we give the transformation of field operator under the Uglov's mapping from multi-component system to single-component one. 
As an application, hole propagator is calculated by using this method. With introducing renormalized momenta and spin variables of quasi-holes, the expression for hole propagator in finite-sized system becomes much simpler than that derived in \cite{Ka2}. We confirm that the expression in the thermodynamic limit recovers the earlier result\cite{Ka2}.  Furthermore we discuss spectral function of hole propagator for a specific coupling constant ($\lambda=1$ in the notations we will introduce in the following chapters). At this coupling parameter, the hole propagator of spin Calogero-Sutherland model is equivalent to the dynamical colour correlation function\cite{Yamamoto2000,Yamamoto1999} of SU(3) Haldane-Shastry model\cite{Ha2,Kaw,kiwata} as shown by Arikawa\cite{Ari}. 
In the same way as hole propagator, particle propagator can be mapped to that of single-component model. However, it is more involved to take the thermodynamic limit of particle propagator than hole propagator. Therefore, we report the calculation of particle propagator and discussion of the corresponding spectral weight in a separate paper. 

To outline this paper,
the basic properties of Spin Calogero-Sutherland model
are shown in section \ref{B}.
In section \ref{A}, the mapping of the field annihilation operator is 
considered.
Using the result of section \ref{A},
hole propagator of spin Calogero-Sutherland model is 
derived in section \ref{H}. The expression for hole propagator is rewritten in terms of quasi-hole rapidities in section \ref{Q}. The properties of the spectral function 
for $\lambda=1$ are discussed in section \ref{S}.

\section{Basic Properties}\label{B}

In this section, we review the basic properties of spin
Calogero-Sutherland model.

\subsection{Hamiltonian and eigenfunctions}

Spin Calogero-Sutherland model 
is a one-dimensional quantum model which
consists of $N$ particles with spin degrees of freedom, 
moving along the circle of perimeter $L$.
Each pair of particles has an interaction of inverse square type potential. 
Hamiltonian is given by
\begin{equation}
 H=-\sum_{i=1}^N\frac{\partial^2}{\partial x_i^2}
 +\frac{2\pi^2}{L^2}\sum_{i<j}\frac{\lambda(\lambda+P_{ij})}
 {\displaystyle \sin^2 \left[\pi(x_i-x_j)/L\right]}, \label{spinCSM}
\end{equation}
where $x_i$ is the coordinate of $i$-th particle, and
$P_{ij}$ is the spin exchange operator for particles $i$ and $j$.
The Hamiltonian (\ref{spinCSM}) has one parameter, 
an interaction parameter $\lambda$, 
which controls whole physical properties of the system.
Physically, $\lambda< 0$ is unrealistic due to the collapse of the particles.
In the following of this paper, we consider particles with spin 1/2,
and $\lambda$ is restricted to be non-negative integer for simplicity. 

Let the particles be bosons for odd $\lambda$ and fermions for even $\lambda$, 
following to 
the earlier works on the hole propagator\cite{Kat,Ka2}, 
and the boundary condition is chosen to be periodic. 
In the ground state, 
$N/2$ is taken to be odd (even) when $\lambda$ is even (odd) 
to avoid the degeneracy of the ground state.
Introducing new variables $z=\{z_1,\cdots,z_N\}$ with 
$z_i=\exp[2\pi \rmi x_i/L]$, 
the wave function $\Psi(\{x_i\},\{\sigma_i\})$ for spin $1/2$ model is written 
as the product $\Phi(z,\sigma)\Psi_{0,N}(z)$ of Jastrow type wave function 
\begin{equation}
\Psi_{0,N}(z)=\prod_{i}z_i^{-\lambda(N-1)/2}\prod_{i<j}(z_i-z_j)^{\lambda}
\label{eq:Psi-zero}
\end{equation}
 and a function $\Phi(z,\sigma)$ of complex spatial coordinates $z$ 
and spin variables $\sigma=\{\sigma_1,\cdots,\sigma_N\}$. 
The spin coordinate $\sigma_i$ takes 1 (2) for spin up (spin down). 

When $\lambda$ is even, particles are fermions 
and the Jastrow wave function $\Psi_{0,N}(z)$ is symmetric 
with respect to interchange of $z_i$ and $z_j$. $\Phi(z,\sigma)$ 
then obeys the fermionic Fock condition, 
\begin{equation}
\Phi(\cdots,z_i,\sigma_i,\cdots,z_j,\sigma_j,\cdots)
=-
\Phi(\cdots,z_j,\sigma_j,\cdots,z_i,\sigma_i,\cdots).
\label{eq:Phi-fock}
\end{equation}
When $\lambda$ is odd, on the other hand, 
the particles are bosons and the Jastrow wave function $\Psi_{0,N}(z)$ is 
anti-symmetric with respect to interchange of $z_i$ and $z_j$. 
Thus $\Phi(z,\sigma)$ obeys the fermionic Fock condition (\ref{eq:Phi-fock}). 

The Jastrow wave function $\Psi_{0,N}$ is periodic under the translation $x_i\rightarrow x_i+L$ when $\lambda$ is even or 
$N$ is odd and anti-periodic otherwise. It thus follows that 
\begin{equation}
\Phi(z,\sigma)\mbox{ is }\left\{
\begin{array}{cc}
\mbox{periodic } & \mbox{when $\lambda$ is even or $N$ is odd}\\ 
\mbox{anti-periodic } & \mbox{otherwise}. 
\end{array}
\right.
\label{eq:pbc-Phi}
\end{equation}
A basis for wave function satisfying 
(\ref{eq:Phi-fock}) and (\ref{eq:pbc-Phi}) is given by 
the Slater determinant of free fermions with spin 1/2 
\begin{equation}
u_{\kappa,\alpha}={\rm Asym}\left[\,\prod_{i=1}^N z^{\kappa_i} \varphi_{\alpha_i}(\sigma_i)\right]
\label{eq:ukappaalpha}
\end{equation}
for a set of momenta $\kappa=(\kappa_1,\cdots,\kappa_N)$ and
 spin configuration $\alpha=(\alpha_1,\cdots,\alpha_N)$.
Here 
$\alpha_i=1,2$ means $z$ component of spin of 
$i$-th particle being $+1/2$ and $-1/2$, respectively.
Here one-particle spin function $\varphi_{\alpha_i}(\sigma_i)$ is given by $\delta_{3/2-\alpha_i,\sigma_i}$. 
The symbol Asym$[\cdots]$ means anti-symmetrization 
of the function of $z$ and $\sigma$
\begin{eqnarray}
&{\rm Asym}\ f(z_1,\sigma_1,\cdots, z_N,\sigma_N) \nonumber\\
=&\sum_{P\in S_N}\left(-1\right)^{P}f(z_{P(1)},\sigma_{P(1)},\cdots, 
z_{P(N)},\sigma_{P(N)}),
\end{eqnarray}
where $\left(-1\right)^{P}$ denotes the sign of the permutation $P$ 
in the symmetric group $S_N$.

When the basis function (\ref{eq:ukappaalpha})
obeys the periodic boundary condition, the set of momenta $\kappa$ belongs to
\begin{equation}
{\cal L}_{N,2}=
\left\{ 
\kappa=(\kappa_1,\kappa_2,\cdots,\kappa_N)\in {\cal L}_N|
\,\forall s\in {\bf{Z}},\sharp\{\kappa_i \: | \: \kappa_i =  s\}  
\leq 2 \right\},
\label{eq:calL-N-2}
\end{equation} 
which is a subset of 
\begin{equation}
{\cal L}_{N}=
\left\{ 
\kappa=(\kappa_1,\cdots,\kappa_N)\in \mbox{{\bf Z}}^N|
\,\kappa_i\ge \kappa_{i+1}\mbox{ for }i\in [1,N-1]\right\}.
\label{eq:calL-N}
\end{equation} 
We call the elements of ${\cal L}_N$ by shifted partitions. 
When the basis function (\ref{eq:ukappaalpha})
obeys the anti-periodic boundary condition, $\kappa$ belongs to ${\cal L}'_{N,2}$, which is defined by 
\begin{equation}
{\cal L}'_{N,2}=
\left\{ 
\kappa\,|\,
\kappa+1/2\equiv(\kappa_1+1/2,\cdots,\kappa_N+1/2)\in {\cal L}_{N,2}\right\}.
\label{eq:calLprime-N}
\end{equation} 

For spin $1/2$ system, 
the definition (\ref{eq:calL-N-2}) comes from the fact 
that a single orbital state can accommodate at most two particles. 
Furthermore a pair of particles with the same momentum $\kappa_i=\kappa_{i+1}$
cannot have the same spin state, i.e., $\alpha_i\neq\alpha_{i+1}$
owing to Pauli exclusion principle.
For a given set of momenta $\kappa\in {\cal L}_{N,2}$ or ${\cal L}'_{N,2}$, 
therefore, each spin configuration is specified 
by the element of $W_\kappa$ defined as
\begin{equation}
W_\kappa=\left\{\alpha=(\alpha_1,\cdots,\alpha_N)\in [1,2]^N|\,\alpha_i 
<\alpha_{i+1} \mbox{  if  }\kappa_i=\kappa_{i+1}\right\}.
\end{equation}
For $N=2$, $W_\kappa$ is given by  
$\{(1,1), (1,2), (2,1), (2,2)\}$ 
when $\kappa_1>\kappa_2$,
and $\{(1,2)\}$
when $\kappa_1=\kappa_2$.
Thus the basis function $u_{\kappa,\alpha}$ is uniquely specified by 
$(\kappa,\alpha)\in ({\cal L}_{N,2},W_\kappa)$ 
under the periodic boundary condition 
and $(\kappa,\alpha)\in ({\cal L}'_{N,2},W_\kappa)$ 
under the anti-periodic boundary condition. 

Now we define the ordering between the basis functions. 
First we introduce dominance partial order\cite{Mac}
\begin{eqnarray}
 \nu\,>\,\mu \quad\Leftrightarrow\quad |\nu|=|\mu|
 \,\,\,\,\, \mbox{and}\,\,\,\,\, 
 \forall r>0 \,\,\,,\,\, \sum_{i=1}^r \nu_i\,>\,
 \sum_{i=1}^r \mu_i
\end{eqnarray}
between $\nu,\mu\in {\cal L}_{N,2}$ or $\nu,\mu\in {\cal L}'_{N,2}$. 
Next we define the order for spin configurations as
\begin{eqnarray}
 \alpha>\alpha'\,\,\,\Leftrightarrow\,\,\,
 &\sum_{i=1}^N\alpha_i=\sum_{i=1}^N\alpha'_i, \,\, \nonumber\\
 &\mbox{and }\mbox{nonzero }
 \alpha'_i-\alpha_i\,
 \mbox{ at the least $i$ is positive}.
\end{eqnarray}
For example, spin configurations with $N=3$ and $\sum_i^N\alpha_i=4$ 
are arranged as
\begin{eqnarray}
 (2,1,1)<(1,2,1)<(1,1,2).
\end{eqnarray}
The order of $(\kappa,\alpha)$ is then defined by
\begin{eqnarray}
 (\kappa,\alpha)>(\kappa',\alpha')\quad\Leftrightarrow\quad
 \,\kappa>\kappa'\, ,
 \,\,\,\,\mbox{or}\quad 
 \kappa=\kappa' \,\,\,\mbox{and}\,\,\, \alpha>\alpha'.
\end{eqnarray} 
Uglov showed \cite{Ugl} that 
the excited parts of the eigenfunctions for the Hamiltonian 
(\ref{spinCSM}) is characterized by $(\kappa,\alpha)$ as 
$\Phi_{\kappa,\alpha}(z,\sigma)$, which can be uniquely defined by the following two conditions:  
\begin{enumerate}
 \item[(i)] 
       $\Phi_{\kappa,\alpha}(z,\sigma)$ is expanded by $u_{\kappa',\alpha'}$
       satisfying $(\kappa',\alpha')\le(\kappa,\alpha)$
\begin{eqnarray}
 \Phi_{\kappa,\alpha}(z,\sigma)=u_{\kappa,\alpha}
 +\sum_{(\kappa',\alpha')(<(\kappa,\alpha))}
 a_{(\kappa',\alpha')(\kappa,\alpha)}u_{\kappa',\alpha'}. \label{YGZcon1}
\end{eqnarray}
 \item[(ii)] Orthogonal with respect to the norm $\langle\cdots\rangle_{N,\lambda}$ 
\begin{eqnarray}
 \langle\Phi_{\kappa',\alpha'},
 \Phi_{\kappa,\alpha}\rangle_{N,\lambda}=0
 \quad \mbox{for} \,\,(\kappa',\alpha')\neq(\kappa,\alpha), \label{YGZcon2}
\end{eqnarray}
\end{enumerate}
where the scalar product $\langle\cdots\rangle_{N,\lambda}$ is defined
by a weighted integral 
\begin{eqnarray}
 &\langle \Phi',\Phi \rangle_{N,\lambda}
 = \nonumber\\
 &\,\,\frac{1}{N!} 
 \left[\prod_{i=1}^N
 \oint\frac{{\rm d}z_i}{2\pi \rmi z_i}\,\sum_{\sigma_i}\right]
 \prod_{i\neq j}\left(1-\frac{z_i}{z_j}\right)^{\lambda}
 \overline{\Phi'(z,\sigma)}\,\Phi(z,\sigma) \label{ynorm}
\end{eqnarray}
($\overline{\Phi(z,\sigma)}$
means complex conjugate of $\Phi(z,\sigma)$).
The scalar product (\ref{ynorm}) comes from the usual one: 
\begin{eqnarray}
&\langle\Psi'|\Psi\rangle\nonumber\\
&=
\sum_{\sigma_1=\pm 1/2}\cdots\sum_{\sigma_N=\pm 1/2}\int_0^L{\rm d}x_1\cdots 
\int_0^L{\rm d} x_N 
\overline{\Psi'(\{x_i\},\{\sigma_i\})}
          \Psi(\{x_i\},\{\sigma_i\})\label{eq:physical-norm}.
\end{eqnarray}
The relation 
\begin{eqnarray}
&\langle\Psi'|\Psi\rangle
=
N! L^N\langle\Phi',\Phi\rangle_{N,\lambda}
\end{eqnarray}
holds when $\Psi=\Phi\Psi_{0,N}$ and $\Psi'=\Phi'\Psi_{0,N}$. 
The eigenenergy of the eigenfunction $\Phi_{\kappa,\alpha}\Psi_{0,N}$ is given by
\begin{equation}
E_{N}(\kappa)=(2\pi /L)^2\sum_i^{N} 
 \left(\kappa_i+\lambda (N+1-2i)/2\right)^2
 \label{eq:ENkappa}.
\end{equation}
For example, the ground state of $N$-particle system is specified by\cite{Tak} 
$\kappa^0,\alpha^0$ with 
\begin{eqnarray}
 \kappa^0&=\left(\frac{N}{4}-\frac{1}{2},\frac{N}{4}-\frac{1}{2},
 \frac{N}{4}-\frac{3}{2},\cdots,
 -\frac{N}{4}+\frac{1}{2},-\frac{N}{4}+\frac{1}{2}\right) 
 \label{eq:kappazero}
\end{eqnarray}
and 
$$\alpha^0=\left(1,2,1,\cdots,1,2\right).$$
The ground state energy $E_{N}({\rm g})$ is then given by\cite{Ha2,Min}\begin{equation}
E_{N}({\rm g})=(2\pi/L)^2[(1+2\lambda)^2 (M^2-1)M/3+\lambda^2 M/2]
\label{eq:groundenergy-N}
\end{equation}
with $M=N/2$. 

The above properties (\ref{YGZcon1}), (\ref{YGZcon2}) 
and (\ref{ynorm}) are used for the mapping of the eigenfunctions of
spin $1/2$ Calogero-Sutherland model in Uglov's method\cite{Ugl}.
The polynomials $\Phi_{\kappa,\alpha}(z,\sigma)$
defined by the above two conditions
(\ref{YGZcon1}) and (\ref{YGZcon2})
are known to coincide with
Yangian Gelfand-Zetlin basis\cite{Ugl,Tak},
while Yangian Gelfand-Zetlin basis for spin 1/2 system
is originally defined as the simultaneous
eigenfunctions of the quantum determinants of Yangian
$Y(\mathfrak{gl}_1)$ and $Y(\mathfrak{gl}_2)$\cite{Ugl,Naz,Tak}.
In the following, therefore, 
we refer to $\Phi_{\kappa,\alpha}(z,\sigma)$
as Yangian Gelfand-Zetlin basis.
\subsection{Macdonald polynomials and gl$_2$-Jack polynomials}
%
%
Macdonald polynomials\cite{Mac} are symmetric polynomials with two parameters. 
 As a limit of Macdonald polynomials, symmetric Jack polynomials and Schur polynomials can be derived. 
A lot of mathematical relations of
 Macdonald polynomial are known and provide important formulas of Jack polynomials, which have been utilized in calculation of correlation functions of scalar Calogero-Sutherland model\cite{Ha94,Ha95a,Ha95b,Les,Ser}. 

Macdonald polynomials themselves are eigenfunctions of Ruijsenaars-Schneider model\cite{Rui,Rui2}, which is a relativistic generalization of 
 scalar Calogero-Sutherland model.
The dynamical correlation functions 
of Ruijsenaars-Schneider model
have been calculated
with use of the properties of Macdonald polynomials\cite{Kon}.
%

As an index which specifies each symmetric polynomial, 
we define partitions as the set of non-negative integers 
arranged in non-increasing order. 
We denote the set of partitions with length equal to or shorter than $N$ by 
\begin{eqnarray}
 \Lambda_N=&\{
 \nu=(\nu_1,\nu_2,\cdots,\nu_N)\in{\bf Z}^N 
 \,| \nu_1\ge\nu_2\ge\cdots\ge\nu_N\ge 0\,\}.
\end{eqnarray}
The monomial symmetric polynomial $m_{\nu}$ with a partition 
$\nu\in \Lambda_N$ is defined by symmetrization of
a monomial $z^{\nu}=z_1^{\nu_1}z_2^{\nu_2}\cdots z_N^{\nu_N}$ as 
\begin{eqnarray}
 m_{\nu}=
 \sum_{\sigma}z_1^{\nu_{\sigma(1)}}z_2^{\nu_{\sigma(2)}}
 \cdots z_N^{\nu_{\sigma(N)}},
\end{eqnarray}
where the sum is taken for all distinct permutations of the elements of $\nu$.

Macdonald polynomial $P_{\nu}(z;q,t)$ for $\nu\in \Lambda_N$ 
is uniquely defined 
by the following two conditions\cite{Mac}:
\begin{enumerate}
 \item[(i)] $P_{\nu}(z;q,t)$ is expanded by $m_{\mu}$
       satisfying $\mu\le \nu$
\begin{eqnarray}
 P_{\nu}(z;q,t)=m_{\nu}+\sum_{\mu(<\nu)}v_{\nu\mu}m_{\mu}. \label{maccon1}
\end{eqnarray}
 \item[(ii)] Orthogonal with respect to the norm $\langle \cdots \rangle_{N,q,t}$
\begin{eqnarray}
 \langle P_{\mu}(z;q,t),P_{\nu}(z;q,t) \rangle_{N,q,t} = 0
 \quad \mbox{for} \,\,\mu\neq\nu, \label{maccon2}
\end{eqnarray}
\end{enumerate}
where the norm of (\ref{maccon2}) is defined by a weighted integral
through the function $(x;q)_{\infty}=\prod_{r=0}^{\infty}(1-xq^r)$
\begin{eqnarray}
 \langle f,g \rangle_{N,q,t} =
 \frac{1}{N!} 
 \left[\prod_{i=1}^N
 \oint\frac{{\rm d}z_i}{2\pi iz_i}\right]
 \prod_{i\neq j}
 \frac{\displaystyle (z_i/z_j;q)_{\infty}}
 {\displaystyle (tz_i/z_j;q)_{\infty}}
 \overline{f(z)}\,g(z). \label{mnorm}
\end{eqnarray}
Since symmetric Jack polynomial is the limit $t=q^{\lambda},q\to 1$
of Macdonald polynomial, the above two conditions
also define symmetric Jack polynomial uniquely by taking the limit 
$t=q^{\lambda},q\to 1$ of the norm (\ref{mnorm}).
Schur symmetric polynomials $ s_{\nu}(z)$ for $\nu\in \Lambda_N$ 
\begin{equation}
 s_{\nu}(z)=\frac{\mbox{Asym}\left[z_1^{\nu_1+N-1}z_2^{\nu_2+N-2}
 \cdots z_N^{\nu_N}\right]}
 {\mbox{Asym}\left[z_1^{N-1}z_2^{N-2}\cdots z_N^{0}\right]}
\end{equation}
can also be obtained as the limit $t=q\rightarrow 1$ of $P_{\nu}(z;q,t)$.
Uglov utilized the properties of Macdonald polynomials
to calculate the dynamical correlation functions
by mapping Yangian Gelfand-Zetlin basis
to symmetric polynomials that are a limit of Macdonald polynomials\cite{Ugl}.
These new polynomials are called gl$_2$-Jack polynomials\cite{Ugl,Yam},
and defined as  
\begin{equation}
P_{\nu}^{(2\lambda+1)}(z)=
\lim_{q=-p,t=-p^{2\lambda+1},p\rightarrow 1} P_{\nu}(z). 
\label{eq:gl2-def}
\end{equation}
In the following, we call the limit in (\ref{eq:gl2-def}) ``Uglov limit''. 
From (\ref{maccon1}), (\ref{maccon2}) and (\ref{eq:gl2-def}), it follows that 
\begin{enumerate}
 \item[(i)] $P_{\nu}^{(2\lambda+1)}(z)$ is expanded by $m_{\mu}$
       satisfying $\mu\le \nu$
\begin{eqnarray}
 P_{\nu}^{(2\lambda+1)}(z)=m_{\nu}+\sum_{\mu(<\nu)}c_{\nu\mu}m_{\mu}.
 \label{gl2con1}
\end{eqnarray}
 \item[(ii)] Orthogonal with respect to the scalar product $\{ \cdots\}_{N,\lambda}$
\begin{eqnarray}
 \{ P_{\mu}^{(2\lambda+1)}
 , P_{\nu}^{(2\lambda+1)}\}_{N,\lambda} = 0
 \quad \mbox{for} \,\,\mu\neq\nu. \label{gl2con2}
\end{eqnarray}
\end{enumerate}
The scalar product in (\ref{gl2con2}) is
defined as
\begin{eqnarray}
 &\left\{f,g\right\}_{N,\lambda}
 = \nonumber\\
 &\,\,\frac{1}{N!} 
 \left[\prod_{i=1}^N
 \oint\frac{{\rm d}z_i}{2\pi iz_i}\right]
 \prod_{i\neq j}\left(1-\frac{z_i}{z_j}\right)^{\lambda+1}\!\!
 \left(1+\frac{z_i}{z_j}\right)^{\lambda}
 \overline{f(z)}g(z),\label{gnorm}
\end{eqnarray}
which comes from a limit of (\ref{mnorm}). 
\ \\
The two properties (\ref{gl2con1}) and (\ref{gl2con2}) can be regarded 
as the defining properties of gl$_2$-Jack polynomials. 
Alternatively, we can define gl$_2$-Jack polynomials by 
\begin{equation}
P_{\nu}^{(2\lambda+1)}(z)=s_{\nu}+\sum_{\mu(<\nu)}C_{\nu\mu}s_{\mu}
 \label{gl2con1prime}
\end{equation}
and (\ref{gl2con2}) because (\ref{gl2con1prime}) and (\ref{gl2con1}) are equivalent as shown below. 
\ \\
Schur polynomials, which is a limit of Macdonald polynomials, 
 can be expanded by monomial symmetric polynomials,
and conversely monomial symmetric polynomials are 
written in the form of 
\begin{eqnarray}
 m_{\nu}(z)=s_{\nu}+\sum_{\nu'(<\nu)}a_{\nu\nu'}s_{\nu'},
 \label{scon1}
\end{eqnarray}
from which
\begin{eqnarray}
 P_{\nu}^{(2\lambda+1)}(z)
 &=m_{\nu}+ \sum_{\mu(<\nu)}c_{\nu\mu}m_{\mu} \nonumber\\
 &=s_{\nu}+ \sum_{\nu'(<\nu)}a_{\nu\nu'}s_{\nu'} +
 \sum_{\mu(<\nu)}c_{\nu\mu}
 \left(s_{\mu}+ \sum_{\mu'(<\mu)}a_{\mu\mu'}s_{\mu'}\right) \nonumber\\
 &=s_{\nu}+\sum_{\mu(<\nu)}C_{\nu\mu}s_{\mu}
\end{eqnarray}
follows. 

So far Macdonald polynomials $P_\nu$ and gl$_2$-Jack polynomials 
$P^{(2\lambda+1)}_\nu$ have been defined for a partition $\nu\in \Lambda_N$. 
However, it is convenient to extend the definition of $P_\nu$ and 
$P^{(2\lambda+1)}_\nu$ for $\nu$ that belongs to ${\cal L}_N$ (or ${\cal L}_N'$).
When $\nu\in {\cal L}_N$ is given by $$\nu=\mu-J=(\mu_1 -J,\cdots,\mu_N-J)$$ 
with an integer (half integer) $J$ and a partition $\mu\in \Lambda_N$, 
we define $P_\nu$ and $P^{(2\lambda+1)}_\nu$ as
\begin{equation}
P_\nu(z)\equiv (z_1 \cdots z_N)^{-J}P_\mu(z),\quad
P^{(2\lambda+1)}_\nu(z)\equiv (z_1 \cdots z_N)^{-J}P^{(2\lambda+1)}_\mu(z),
\end{equation}
respectively. 
\subsection{Mapping from Yangian Gelfand-Zetlin basis 
to gl$_2$-Jack polynomials}
\begin{figure}[t]
 \begin{center}
\unitlength 0.1in
\begin{picture}( 48.9000,  9.9000)(  3.2000,-15.9000)
%
\special{pn 8}%
\special{pa 1400 600}%
\special{pa 1580 600}%
\special{pa 1580 780}%
\special{pa 1400 780}%
\special{pa 1400 600}%
\special{fp}%
%
\special{pn 8}%
\special{pa 1580 780}%
\special{pa 1760 780}%
\special{pa 1760 960}%
\special{pa 1580 960}%
\special{pa 1580 780}%
\special{fp}%
%
\special{pn 8}%
\special{pa 1400 780}%
\special{pa 1580 780}%
\special{pa 1580 960}%
\special{pa 1400 960}%
\special{pa 1400 780}%
\special{fp}%
%
\special{pn 8}%
\special{pa 1400 960}%
\special{pa 1580 960}%
\special{pa 1580 1140}%
\special{pa 1400 1140}%
\special{pa 1400 960}%
\special{fp}%
%
\special{pn 8}%
\special{pa 1580 960}%
\special{pa 1760 960}%
\special{pa 1760 1140}%
\special{pa 1580 1140}%
\special{pa 1580 960}%
\special{fp}%
%
\special{pn 8}%
\special{pa 1580 1140}%
\special{pa 1760 1140}%
\special{pa 1760 1320}%
\special{pa 1580 1320}%
\special{pa 1580 1140}%
\special{fp}%
%
\special{pn 8}%
\special{pa 1400 1140}%
\special{pa 1580 1140}%
\special{pa 1580 1320}%
\special{pa 1400 1320}%
\special{pa 1400 1140}%
\special{fp}%
%
\special{pn 8}%
\special{pa 320 1140}%
\special{pa 500 1140}%
\special{pa 500 1320}%
\special{pa 320 1320}%
\special{pa 320 1140}%
\special{fp}%
%
\special{pn 8}%
\special{pa 680 780}%
\special{pa 860 780}%
\special{pa 860 960}%
\special{pa 680 960}%
\special{pa 680 780}%
\special{fp}%
%
\special{pn 8}%
\special{pa 680 600}%
\special{pa 860 600}%
\special{pa 860 780}%
\special{pa 680 780}%
\special{pa 680 600}%
\special{fp}%
%
\special{pn 8}%
\special{pa 500 1140}%
\special{pa 680 1140}%
\special{pa 680 1320}%
\special{pa 500 1320}%
\special{pa 500 1140}%
\special{fp}%
%
\special{pn 8}%
\special{pa 680 960}%
\special{pa 680 1140}%
\special{fp}%
%
\special{pn 8}%
\special{pa 2800 1140}%
\special{pa 2980 1140}%
\special{pa 2980 1320}%
\special{pa 2800 1320}%
\special{pa 2800 1140}%
\special{fp}%
%
\special{pn 8}%
\special{pa 2980 1140}%
\special{pa 3160 1140}%
\special{pa 3160 1320}%
\special{pa 2980 1320}%
\special{pa 2980 1140}%
\special{fp}%
%
\special{pn 8}%
\special{pa 3160 1140}%
\special{pa 3340 1140}%
\special{pa 3340 1320}%
\special{pa 3160 1320}%
\special{pa 3160 1140}%
\special{fp}%
%
\special{pn 8}%
\special{pa 3340 1140}%
\special{pa 3520 1140}%
\special{pa 3520 1320}%
\special{pa 3340 1320}%
\special{pa 3340 1140}%
\special{fp}%
%
\special{pn 8}%
\special{pa 3520 1140}%
\special{pa 3700 1140}%
\special{pa 3700 1320}%
\special{pa 3520 1320}%
\special{pa 3520 1140}%
\special{fp}%
%
\special{pn 8}%
\special{pa 3700 1140}%
\special{pa 3880 1140}%
\special{pa 3880 1320}%
\special{pa 3700 1320}%
\special{pa 3700 1140}%
\special{fp}%
%
\special{pn 8}%
\special{pa 2800 960}%
\special{pa 2980 960}%
\special{pa 2980 1140}%
\special{pa 2800 1140}%
\special{pa 2800 960}%
\special{fp}%
%
\special{pn 8}%
\special{pa 2980 960}%
\special{pa 3160 960}%
\special{pa 3160 1140}%
\special{pa 2980 1140}%
\special{pa 2980 960}%
\special{fp}%
%
\special{pn 8}%
\special{pa 2620 600}%
\special{pa 2800 600}%
\special{pa 2800 780}%
\special{pa 2620 780}%
\special{pa 2620 600}%
\special{fp}%
%
\special{pn 8}%
\special{pa 2800 960}%
\special{pa 2800 780}%
\special{fp}%
%
\special{pn 8}%
\special{pa 4670 600}%
\special{pa 4850 600}%
\special{pa 4850 780}%
\special{pa 4670 780}%
\special{pa 4670 600}%
\special{fp}%
%
\special{pn 8}%
\special{pa 4850 600}%
\special{pa 5030 600}%
\special{pa 5030 780}%
\special{pa 4850 780}%
\special{pa 4850 600}%
\special{fp}%
%
\special{pn 8}%
\special{pa 5030 600}%
\special{pa 5210 600}%
\special{pa 5210 780}%
\special{pa 5030 780}%
\special{pa 5030 600}%
\special{fp}%
%
\special{pn 8}%
\special{pa 4490 960}%
\special{pa 4670 960}%
\special{pa 4670 1140}%
\special{pa 4490 1140}%
\special{pa 4490 960}%
\special{fp}%
%
\special{pn 8}%
\special{pa 4490 1140}%
\special{pa 4670 1140}%
\special{pa 4670 1320}%
\special{pa 4490 1320}%
\special{pa 4490 1140}%
\special{fp}%
%
\special{pn 8}%
\special{pa 4670 960}%
\special{pa 4670 780}%
\special{fp}%
\put(3.3000,-15.2000){\makebox(0,0)[lb]{$\kappa=$}}%
\put(27.6000,-15.3000){\makebox(0,0)[lb]{$\alpha-2\kappa=$}}%
\put(45.6000,-15.4000){\makebox(0,0)[lb]{$\nu=$}}%
\put(13.2000,-15.2000){\makebox(0,0)[lb]{$\alpha=$}}%
\put(13.1000,-17.6000){\makebox(0,0)[lb]{$(1,2,2,2)$}}%
\put(3.3000,-17.6000){\makebox(0,0)[lb]{$(1,1,0,-2)$}}%
\put(27.6000,-17.6000){\makebox(0,0)[lb]{$(-1,0,2,6)$}}%
\put(45.6000,-17.6000){\makebox(0,0)[lb]{$(3,0,-1,-1)$}}%
%
\special{pn 8}%
\special{pa 2100 960}%
\special{pa 2400 960}%
\special{fp}%
\special{sh 1}%
\special{pa 2400 960}%
\special{pa 2334 940}%
\special{pa 2348 960}%
\special{pa 2334 980}%
\special{pa 2400 960}%
\special{fp}%
%
\special{pn 8}%
\special{pa 4000 960}%
\special{pa 4300 960}%
\special{fp}%
\special{sh 1}%
\special{pa 4300 960}%
\special{pa 4234 940}%
\special{pa 4248 960}%
\special{pa 4234 980}%
\special{pa 4300 960}%
\special{fp}%
\put(10.9000,-10.5000){\makebox(0,0)[lb]{,}}%
\end{picture}%
 \end{center}
 \caption{An example of the transformation (\ref{parttrans}) for $N=4$.
 While $(\kappa,\alpha)\in ({\cal L}_{N,2},W_\kappa)$ or $({\cal L}'_{N,2},W_\kappa)$, $\alpha-2\kappa$ is a shifted partition of the spinless fermionic system in reverse order,
 and $\nu$ is a shifted partition of the spinless bosonic system.}
 \label{trans}
\end{figure}
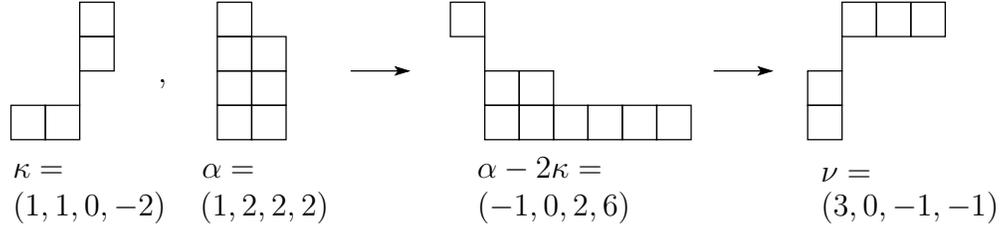

Uglov defined\cite{Ugl} the linear mapping $\Omega$ between 
the set of functions spanned by $u_{\kappa,\alpha}$ with $(\kappa,\alpha)\in ({\cal L}_{N,2},W_\kappa)$ or $({\cal L}'_{N,2},W_\kappa)$ and the set of symmetric functions as 
$\Omega(u_{\kappa,\alpha})=s_{\nu}$,
where the relation between $(\kappa,\alpha)$ and $\nu$ is given by
\begin{eqnarray}
 \nu_i=\alpha_{N+1-i}-2\kappa_{N+1-i}-N+i \label{parttrans}
\end{eqnarray}
for the system of $N$ particles with spin $1/2$.
An example of this transformation is drawn in Figure \ref{trans}.
The properties of $\Omega$ are listed as follows:
\begin{enumerate}
\item[(i)] Isometry. The scalar product is preserved under the mapping $\Omega$. 
For functions $\Phi'(z,\sigma)$ and $\Phi(z,\sigma)$, the relation 
\begin{equation}
\langle \Phi',\Phi\rangle_{N,\lambda}=
\left\{\Omega(\Phi'),\Omega(\Phi)\right\}_{N,\lambda}
\end{equation}
holds.
\item[(ii)]
For any symmetric function $f(z)$ \cite{Ugl}, the relation
\begin{eqnarray}
 \Omega(f(z_1,\cdots,z_{N})\,u_{\kappa,\alpha}(z,\sigma))
 =f(z_1^{-2},\cdots,z_{N}^{-2})\,\Omega(u_{\kappa,\alpha}(z,\sigma))
\label{eq:gl2-property}
\end{eqnarray}
holds. 
\item[(iii)]
The correspondence between Yangian Gelfand-Zetlin basis and gl$_2$-Jack polynomials 
\begin{equation}
\Omega(\Phi_{\kappa,\alpha})=P^{(2\lambda+1)}_\nu.
\end{equation}
\end{enumerate}
Taking the mapping $\Omega$ for 
both sides of (\ref{YGZcon1}) and (\ref{YGZcon2}) 
in the conditions that specify Yangian Gelfand-Zetlin basis,
there appear the defining relations of gl$_2$-Jack polynomial 
$P_{\nu}^{(2\lambda+1)}(z)$ (\ref{gl2con1prime}) and (\ref{gl2con2}),
and consequently, the property (iii) follows.
The mapping $\Omega$ can be interpreted as a transformation
from a multi-component system to a single-component system.


\section{Transformation of the Field Annihilation Operator by the Mapping}\label{A}

In this section, we consider the mapping of the field annihilation operator.
In calculating density correlation function 
and spin correlation function,
spin operator and density operator are expressed as 
power sum polynomials, that is, c-numbers\cite{Ugl}. 
Annihilation and creation operators in 1-particle Green's function,
however, are not the case.
We first make sure that 
the sum of the annihilation operators of spin $\uparrow$ and $\downarrow$
on the multi-component model
are mapped to the annihilation operator on single-component model.

\subsection{Action of annihilation operator}

Generally, the action of the field
annihilation operator on a wave function is
implemented by 
fixing the coordinate of one of the particles in the wave function
to that of the annihilation operator as
\begin{eqnarray}
 \psi(x)\Psi(x_1,\cdots,x_{N-1},x_N)
 =\sqrt{N} \zeta^{N-1}\Psi(x_1,\cdots,x_{N-1},x)
\end{eqnarray}
for single-component model, where $\zeta=1$ for boson and 
$\zeta=-1$ for fermion.
Moreover for multi-component model,
the action of the field annihilation operator with spatial coordinate $x$
and spin coordinate $\sigma(=1/2,-1/2)$ is given by
\begin{eqnarray}
& \psi_{\sigma}(x)\Psi(x_1,\sigma_1,\cdots,x_{N-1},\sigma_{N-1},x_N,\sigma_N)\nonumber\\
& =\sqrt{N} \zeta^{N-1}\Psi(x_1,\sigma_1,\cdots,x_{N-1},\sigma_{N-1},x,\sigma).
\end{eqnarray}
%
%
In the following, we write $\psi_\uparrow(x)$ ($\psi_\downarrow(x)$) instead of $\psi_{1/2}(x)$ ($\psi_{-1/2}(x)$) for notational convenience. 
First, we consider the action of 
$(\psi_{\uparrow}(0,0)+\psi_{\downarrow}(0,0))$
on the wave function of spin Calogero-Sutherland model
\begin{eqnarray}
\Psi(x,\sigma)
 =\Phi(z,\sigma)\Psi_{0,N}(z), \label{YGZwf}
\end{eqnarray}
where $\Psi_{0,N}(z)$ is the Jastrow type wave function defined in (\ref{eq:Psi-zero}). 
We can take the action of one of the annihilation operators
by restricting the spin configuration
of the Yangian Gelfand-Zetlin basis of both sides of annihilation operator
when calculating the matrix element.
The sum of the field operators acts on the wave function (\ref{YGZwf}) as
\begin{eqnarray}
 &\!\!\!\!\!
 (\psi_{\uparrow}(0,0)+\psi_{\downarrow}(0,0))\,
 \Psi\nonumber\\
 &=\zeta^{N-1}\sqrt{N} \sum_{\sigma_N=1,2}
 \Phi(z_1,\sigma_1,
 \cdots,z_{N-1},\sigma_{N-1},z_N=1,\sigma_N)\nonumber\\
 &\qquad\qquad\qquad\qquad\qquad\times
 \Psi_{0,N}(z_1,\cdots,z_{N-1},z_N=1) \nonumber\\
 &=\zeta^{N-1}\sqrt{N} \sum_{\sigma_N=1,2}
 \Phi(z_1,\sigma_1,
 \cdots,z_{N-1},\sigma_{N-1},z_N=1,\sigma_N) \nonumber\\
 &\qquad\qquad\qquad\qquad\qquad\times
 \prod_{i=1}^{N-1}z_i^{-\lambda/2}(z_i-1)^{\lambda}
 \Psi_{0,N-1}(z_1,\cdots,z_{N-1}).\label{action-psi-s}
\end{eqnarray}
Next, we consider the action of the field annihilation operator on the wave function
\begin{equation}
 f(z)\tilde{\Psi}_{0,N}(z)
 \label{eq:fPsi}
\end{equation}
where $f(z)$ is a symmetric function of $z$ 
and $\tilde{\Psi}_{0,N}$ is given by
\begin{eqnarray}
 \tilde{\Psi}_{0,N}(z)=
 \prod_{i=1}^N z_i^{-(2\lambda+1)(N-1)/2}
 \prod_{i<j}(z_i-z_j)^{\lambda+1}(z_i+z_j)^{\lambda}.
\end{eqnarray}
This function is the Uglov limit ($q=-p,t=-p^{2\lambda+1},p\rightarrow 1$) of the ground state wave function
for Ruijsenaars-Schneider model\cite{Rui,Rui2}.
Acting the field annihilation operator on (\ref{eq:fPsi}), we obtain
\begin{eqnarray}
 &\!\!\!\!\!
 \psi(0,0)f(z)\tilde{\Psi}_{0,N}(z) \nonumber\\
 &=\zeta^{N-1}\sqrt{N}f(z_1,\cdots,z_{N-1},z_N=1)
 \tilde{\Psi}_{0,N}(z_1,\cdots,z_{N-1},z_N=1) \nonumber\\
 &=\zeta^{N-1}\sqrt{N}f(z_1,\cdots,z_{N-1},z_N=1) \nonumber\\
 &\qquad\qquad\times
 \prod_{i=1}^{N-1} z_i^{-\lambda-1/2}
 (z_i-1)^{\lambda+1}(z_i+1)^{\lambda}
 \tilde{\Psi}_{0,N-1}(z_1,\cdots,z_{N-1}). \label{action-psi}
\end{eqnarray}
We introduce $\tilde{\psi}_s(0,0)$ and $\tilde{\psi}(0,0)$ 
through the relations:
\begin{eqnarray}
 &\tilde{\psi}_{s}(0,0)\Phi\equiv
 (\Psi_{0,N-1})^{-1}\,\,\psi_{s}(0,0)\,\,\Phi\Psi_{0,N},
 \quad s=\uparrow\mbox{or}\downarrow\label{tildepsi-s}\\
 &\tilde{\psi}(0,0)f\equiv
 (\tilde{\Psi}_{0,N-1})^{-1}\,\,\psi(0,0)\,\,f\tilde{\Psi}_{0,N}
 \label{tildepsi}.
\end{eqnarray}
In terms of (\ref{tildepsi-s}) and (\ref{tildepsi}), 
(\ref{action-psi-s}) and (\ref{action-psi}) are rewritten as
\begin{eqnarray}
&(\tilde{\psi}_{\uparrow}(0,0)+\tilde{\psi}_{\downarrow}(0,0))\,
 \Phi\nonumber\\
 &=\zeta^{N-1}\sqrt{N} \sum_{\sigma_N=1,2}
 \Phi(z_1,\sigma_1,
 \cdots,z_{N-1},\sigma_{N-1},z_N=1,\sigma_N)
 \prod_{i=1}^{N-1}z_i^{-\lambda/2}(z_i-1)^{\lambda}\nonumber\\
\label{action-tilde-psi-s}
\end{eqnarray}
and 
\begin{eqnarray}
 &\!\!\!\!\!
 \tilde{\psi}(0,0)f(z)\nonumber\\
 &=\zeta^{N-1}\sqrt{N}f(z_1,\cdots,z_{N-1},z_N=1) 
 \prod_{i=1}^{N-1} z_i^{-\lambda-1/2}
 (z_i-1)^{\lambda+1}(z_i+1)^{\lambda}, \label{action-tilde-psi}
\end{eqnarray}
respectively.

\subsection{Transformation of the field annihilation operator}
In this subsection, we show that 
\begin{eqnarray}
 \Omega\left((\tilde{\psi}_{\uparrow}(0,0)+\tilde{\psi}_{\downarrow}(0,0))
 \Phi
 \right)
 &=(-1)^{(N-1)\lambda}\left(\prod_{i=1}^{N-1}z_i^{1/2}\right)
 \tilde{\psi}(0,0)\Omega(\Phi).\label{eq:tilde-psi-trasf}
\end{eqnarray}
To derive (\ref{eq:tilde-psi-trasf}), it suffices to show 
\begin{eqnarray}
 \Omega\left((\tilde{\psi}_{\uparrow}(0,0)+\tilde{\psi}_{\downarrow}(0,0))
 u_{\kappa,\alpha}
 \right)
 & =(-1)^{(N-1)\lambda}\left(\prod_{i=1}^{N-1}z_i^{1/2}\right)
 \tilde{\psi}(0,0)s_\nu\label{eq:tilde-psi-trasf-u-s}
\end{eqnarray}
with (\ref{parttrans}).
This is because (i) $\Omega$ is a linear operator,  
(ii) $u_{\kappa,\alpha}$ is a basis of $\Phi$ and  
(iii) $\Omega(u_{\kappa,\alpha})=s_\nu$. 

Since $u_{\kappa,\alpha}$ is a Slater determinant 
of $N$ free fermions with spin $1/2$, 
$u_{\kappa,\alpha}$ with one of the coordinate fixed
can be expanded by Slater determinants of 
$N-1$ free fermions with spin $1/2$ as
\begin{eqnarray}
 &\!\!\!\!\!
 \sum_{\sigma=1,2}u_{\kappa,\alpha}(z_1,\sigma_1,\cdots,z_N=1,\sigma)
 \nonumber\\
 &=\sum_{\sigma=1,2}
\left|\matrix{
  z_1^{\kappa_1}\varphi_{\alpha_1}(\sigma_1) & \ldots 
  &z_1^{\kappa_N}\varphi_{\alpha_N}(\sigma_1) \cr
  \vdots & \ddots & \vdots \cr
  z_{N-1}^{\kappa_1}\varphi_{\alpha_1}(\sigma_{N-1}) & \ldots 
  &z_{N-1}^{\kappa_N}\varphi_{\alpha_N}(\sigma_{N-1}) \cr
  \delta_{\alpha_1\sigma} & \ldots & \delta_{\alpha_N\sigma}}\right|\nonumber\\
 &=\sum_{i=1}^N(-1)^{i}
 u_{\cdots,\kappa_{i-1},\kappa_{i+1},\cdots,
 \cdots,\alpha_{i-1},\alpha_{i+1},\cdots}
fs (z_1,\sigma_1,\cdots,z_{N-1},\sigma_{N-1}).
\end{eqnarray}
Therefore,
\begin{eqnarray}
 &\!\!\!\!\!
 \Omega\left((\tilde{\psi}_{\uparrow}(0,0)+\tilde{\psi}_{\downarrow}(0,0))
 u_{\kappa,\alpha}
 \right)
 \nonumber\\
 &=\Omega\bigg(
 \zeta^{N-1}\sqrt{N} \sum_{\sigma_N=1,2}
 u_{\kappa,\alpha}(\{z_i,\sigma_i\})|_{z_N=1}
 \prod_{i=1}^{N-1}z_i^{-\lambda/2}(z_i-1)^{\lambda}\bigg) \nonumber\\
 &=\zeta^{N-1}\sqrt{N}
 \prod_{i=1}^{N-1}z_i^{\lambda}(z_i^{-2}-1)^{\lambda}
 \sum_{i=1}^N(-1)^{i}\nonumber\\
 &\qquad\qquad\times
 \Omega\left(u_{\cdots,\kappa_{i-1},\kappa_{i+1},\cdots,
 \cdots,\alpha_{i-1},\alpha_{i+1},\cdots}
 (z_1,\sigma_1,\cdots,z_{N-1},\sigma_{N-1})\right)  \nonumber
 \\
 &=\zeta^{N-1}\sqrt{N}
 \prod_{i=1}^{N-1}z_i^{\lambda}(z_i^{-2}-1)^{\lambda}
 \sum_{i=1}^N(-1)^{i}\nonumber\\
 &\qquad\qquad\times
 s_{\nu_1+1,\cdots,\nu_{N-i}+1,\nu_{N-i+2},\cdots,\nu_N}
 (z_1,\cdots,z_{N-1}). \label{cut1}
\end{eqnarray}
Here we use the property (\ref{eq:gl2-property}) of $\Omega$. 
We note that though $\nu$ is defined by
$\nu_i=\alpha_{N+1-i}-2\kappa_{N+1-i}-N+i$ for $N$-particle system,
the partition for Schur polynomial on the right-hand side of (\ref{cut1}) 
has $N-1$ elements.

Since Schur polynomial is defined by  Slater determinant of
spinless fermion, the right-hand side of (\ref{cut1}) can be described by 
a Schur polynomial of $N$ variables with one of the variables fixed as
\begin{eqnarray}
 s_{\nu}&(z_1,\cdots,z_{N-1},z_N=1)\nonumber\\
 &=\frac{\mbox{Asym}\left[z_1^{\nu_1+N-1}\cdots z_{N-1}^{\nu_{N-1}+1}1^{\nu_N}
 \right]}
 {\mbox{Asym}\left[z_1^{N-1}\cdots z_{N-1}^{1} 1^{0}\right]}\nonumber\\
 &=\left|\matrix{
  z_1^{\nu_1+N-1} & \ldots &  z_1^{\nu_N} \cr
  \vdots & \ddots & \vdots \cr
  z_{N-1}^{\nu_1+N-1} & \ldots &z_{N-1}^{\nu_N}\cr
  1 & \ldots & 1}\right|
 \cdot\left[\prod_{i=1}^{N-1}(z_i-1)\prod_{1\le i<j\le N-1}(z_i-z_j)\right]^{-1}
 \nonumber\\
 &=\frac{\sum_{i=1}^N (-1)^i \mbox{Asym}
 \left[\cdots z_{i-1}^{\nu_{i-1}+N-(i-1)}
 z_{i+1}^{\nu_{i+1}+N-(i+1)}\cdots\right]}
 {\mbox{Asym}\left[z_1^{N-2}\cdots z_{N-1}^{0}\right]}
 \prod_{i=1}^{N-1}(z_i-1)^{-1}\nonumber\\
 &=\sum_{i=1}^N (-1)^i
 s_{\nu_1+1,\cdots,\nu_{N-i}+1,\nu_{N-i+2},\cdots,\nu_N}(z_1,\cdots,z_{N-1})
 \prod_{i=1}^{N-1}(z_i-1)^{-1}.
\end{eqnarray}
Thus the right-hand side of (\ref{cut1}) is rewritten as
\begin{eqnarray}
 &\mbox{r.h.s. of }(55)\nonumber\\
 &=\zeta^{N-1}
 (-1)^{(N-1)\lambda}\sqrt{N}
 \prod_{i=1}^{N-1}z_i^{-\lambda}(z_i-1)^{\lambda+1}
 (z_i+1)^{\lambda}
 s_{\nu}(z_1,\cdots,z_{N-1},z_N=1) \nonumber\\
 &=(-1)^{(N-1)\lambda}\left(\prod_{i=1}^{N-1}z_i^{1/2}\right)
\tilde{\psi}(0,0)s_{\nu}(z_1,\cdots,z_N)
 .\label{eq:tilde-psi-action}
\end{eqnarray}
From (\ref{cut1}) and (\ref{eq:tilde-psi-action}), the relation 
(\ref{eq:tilde-psi-trasf-u-s}) follows. 
Using (\ref{eq:tilde-psi-trasf}), the matrix element of 
field annihilation operator
is obtained as
\begin{eqnarray}
 &\!\!\!\!\!
 \langle \Phi_{\kappa',\alpha'},
 (\tilde{\psi}_{\uparrow}(0,0)+\tilde{\psi}_{\downarrow}(0,0))
 \Phi_{\kappa,\alpha}
 \rangle_{N-1,\lambda} \nonumber\\
 &=\{\Omega(\Phi_{\kappa',\alpha'}),
 \Omega(
 (\tilde{\psi}_{\uparrow}(0,0)+\tilde{\psi}_{\downarrow}(0,0))
 \Phi_{\kappa,\alpha})
 \}_{N-1,\lambda} \nonumber\\
 &=(-1)^{(N-1)\lambda}
 \left\{P_{\nu'}^{(2\lambda+1)},
 \left(\prod_{i=1}^{N-1}z_i^{1/2}\right)
 \tilde{\psi}(0,0)P_{\nu}^{(2\lambda+1)}\right\}_{N-1,\lambda} \nonumber\\
 &=(-1)^{(N-1)\lambda}
 \left\{P_{\nu'-1/2}^{(2\lambda+1)},
 \tilde{\psi}(0,0)P_{\nu}^{(2\lambda+1)}
 \right\}_{N-1,\lambda}. \label{res2_1}
\end{eqnarray}
Here we note that the relation between $\nu'$ and $(\kappa',\alpha')$
is defined by (\ref{parttrans}) with the number of particles being $N-1$
while the relation between $\nu$ and $(\kappa,\alpha)$ is that with $N$.%
%

 
\section{Combinatorial description of hole propagator}
\label{H}
Hole propagator is one of the 1-particle Green's function defined as
\begin{eqnarray}
 G^-(x,t)&=
 \frac{\langle {\rm g},N|
 \,\psi_{\downarrow}^{\dagger}(x,t)\psi_{\downarrow}(0,0)\,
 |{\rm g},N\rangle}
 {\langle {\rm g},N|{\rm g},N\rangle}. \label{holedef}
\end{eqnarray}
Here $|{\rm g},N\rangle$ is the state vector of $N$-particle ground state, 
whose wave function is given by $\Phi_{\kappa^0,\alpha^0}\Psi_{0,N}$. 
The scalar product in (\ref{holedef}) is the conventional one 
(see (\ref{eq:physical-norm})).

We rewrite (\ref{holedef}) in terms of gl$_2$-Jack polynomial. 
First of all, a complete set of the state vectors is inserted
between the creation operator and the annihilation operator
in the numerator of the right-hand side of (\ref{holedef}).
Next the complex conjugate is taken so as to alter  
the creation operator to the annihilation operator.
Third, we use the relation 
\begin{equation}
\langle(\kappa,\alpha),N-1|\tilde{\psi}_\downarrow(0,0)|{\rm g},N\rangle=
\langle(\kappa,\alpha),N-1|(\tilde{\psi}_\uparrow(0,0)+\tilde{\psi}_\downarrow(0,0))|{\rm g},N\rangle
\end{equation}
when $z$ component of the total spin $S_{z}$ of the state $(\kappa,\alpha)$ is larger than that of the ground state by $1/2$.
The symbol $|(\kappa,\alpha),N-1\rangle$ denotes the state vector of $(N-1)$-particle state whose wavefunction is $\Phi_{\kappa,\alpha}\Psi_{0,N-1}$. 
The scalar product between the state vectors is then 
represented by the scalar product between functions of Yangian Gelfand-Zetlin basis.
Finally we act the mapping $\Omega$ on the wave function and obtain
\begin{eqnarray}
 &G^-(x,t) \nonumber\\
 &=
 \sum_{
  \kappa,\alpha\atop
  {\scriptscriptstyle {\rm s.t.} S^z_{\rm tot}=+1/2}}
 e^{-\rmi\omega_{\kappa}t+\rmi P_{\kappa}x} 
 \frac{|\langle(\kappa,\alpha),N-1|
 (\tilde{\psi}_{\downarrow}(0,0)+\tilde{\psi}_{\uparrow}(0,0))|
 {\rm g},N\rangle|^2}
 {\langle(\kappa,\alpha),N-1\,| \,(\kappa,\alpha),N-1\rangle
 \cdot\langle{\rm g},N\,|\,{\rm g},N\rangle}
 \nonumber\\
 &=\frac{1}{LN}
 \!\!\!
 \sum_{
  \kappa,\alpha\atop
  {\scriptscriptstyle {\rm s.t.} S^z_{\rm tot}=+1/2}}
 \!\!\!
 e^{-\rmi\omega_{\kappa}t+\rmi P_{\kappa}x} 
 \frac{|\langle\Phi_{\kappa,\alpha},
 (\tilde{\psi}_{\downarrow}(0,0)+\tilde{\psi}_{\uparrow}(0,0))
 \Phi_{\mbox{g}}\rangle_{N-1,\lambda}|^2}
 {\langle\Phi_{\kappa,\alpha}\,,\,\Phi_{\kappa,\alpha}\rangle_{N-1,\lambda}
 \langle\Phi_{\mbox{g}}\,,\,\Phi_{\mbox{g}}\rangle_{N,\lambda}}
 \nonumber\\
 &=\frac{1}{LN}
 \!\!\!
 \sum_{
  \nu\in\mathcal{L}_{N-1}\atop
  {\scriptsize {\rm s.t.} S^z_{\rm tot}=+1/2}}
 \!\!\!
 e^{-{\rm i}\omega_{\kappa}t+{\rm i}P_{\kappa}x}
 \frac{|\{P_{\nu-1/2}^{(2\lambda+1)}
 ,\tilde{\psi}(0,0)P_{\mbox{g}}^{(2\lambda+1)}\}_{N-1,\lambda}|^2}
 {\{P_{\nu}^{(2\lambda+1)}, P_{\nu}^{(2\lambda+1)}\}_{N-1,\lambda}
 \{P_{\mbox{g}}^{(2\lambda+1)}, P_{\mbox{g}}^{(2\lambda+1)}\}_{N,\lambda}}.\nonumber\\
 \label{hole}
\end{eqnarray}
The variables of summation $\nu$ is related with $(\kappa,\alpha)$ via
\begin{equation}
\nu_i=\alpha_{N-i}-2\kappa_{N-i}-N+1-i ,\quad i\in [1,N-1].
\end{equation}
The energy $\omega_{\kappa}$ and momentum $P_{\kappa}$ are described
in terms of $\kappa\in {\cal L}_{N,2}$ or ${\cal L}'_{N,2}$ as
\begin{eqnarray}
 &\left(
 \begin{array}{l}
  \displaystyle\omega_{\kappa}=
  E_{N-1}(\kappa)-E_{N}(\mbox{g}) \\[+7pt]
  \displaystyle P_{\kappa}=\frac{2\pi}{L}\sum_i \kappa_i
 \end{array}
 \right. \label{eq:omegakappa}
 \end{eqnarray}
with 
\begin{equation}
E_{N-1}(\kappa)=\left(\frac{2\pi}{L}\right)^2\sum_i^{N-1} 
 \left(\kappa_i+\frac{\lambda (N-2i)}{2}\right)^2\label{eq:EN-1kappa}
\end{equation}
and (\ref{eq:groundenergy-N}). 
$\Omega$ maps $\Phi_{\kappa^0,\alpha^0}$ to 
$P^{(2\lambda+1)}_{\nu^0}\equiv P_{\mbox{g}}^{(2\lambda+1)}$ with 
\begin{equation}
 \nu^{0}=\left(-\frac{N}{2}+2,-\frac{N}{2}+2,\cdots,-\frac{N}{2}+2\right),
\end{equation}
and we obtain gl$_2$-Jack polynomial of the ground state as
$P_{\mbox{g}}^{(2\lambda+1)}(z)=\prod_i z_i^{-N/2+2}$.
The restriction on the sum is considered later.

\subsection{Expansion by gl$_2$-Jack polynomials}
\begin{figure}[t]
 \begin{center}
\unitlength 0.1in
\begin{picture}( 30.9000, 16.9000)(  1.2000,-22.9000)
\put(1.2000,-13.1000){\makebox(0,0)[lb]{$\kappa=(5,5,4,4,4,3,1,1)\,\,\, \Leftrightarrow   $}}%
\put(32.1000,-17.8000){\makebox(0,0)[lb]{$s=(4,3)$}}%
%
\special{pn 8}%
\special{pa 2200 600}%
\special{pa 2398 600}%
\special{pa 2398 798}%
\special{pa 2200 798}%
\special{pa 2200 600}%
\special{fp}%
%
\special{pn 8}%
\special{pa 2398 600}%
\special{pa 2596 600}%
\special{pa 2596 798}%
\special{pa 2398 798}%
\special{pa 2398 600}%
\special{fp}%
%
\special{pn 8}%
\special{pa 2596 600}%
\special{pa 2794 600}%
\special{pa 2794 798}%
\special{pa 2596 798}%
\special{pa 2596 600}%
\special{fp}%
%
\special{pn 8}%
\special{pa 2794 600}%
\special{pa 2992 600}%
\special{pa 2992 798}%
\special{pa 2794 798}%
\special{pa 2794 600}%
\special{fp}%
%
\special{pn 8}%
\special{pa 2992 600}%
\special{pa 3190 600}%
\special{pa 3190 798}%
\special{pa 2992 798}%
\special{pa 2992 600}%
\special{fp}%
%
\special{pn 8}%
\special{pa 2992 798}%
\special{pa 3190 798}%
\special{pa 3190 996}%
\special{pa 2992 996}%
\special{pa 2992 798}%
\special{fp}%
%
\special{pn 8}%
\special{pa 2794 798}%
\special{pa 2992 798}%
\special{pa 2992 996}%
\special{pa 2794 996}%
\special{pa 2794 798}%
\special{fp}%
%
\special{pn 8}%
\special{pa 2596 798}%
\special{pa 2794 798}%
\special{pa 2794 996}%
\special{pa 2596 996}%
\special{pa 2596 798}%
\special{fp}%
%
\special{pn 8}%
\special{pa 2398 798}%
\special{pa 2596 798}%
\special{pa 2596 996}%
\special{pa 2398 996}%
\special{pa 2398 798}%
\special{fp}%
%
\special{pn 8}%
\special{pa 2200 798}%
\special{pa 2398 798}%
\special{pa 2398 996}%
\special{pa 2200 996}%
\special{pa 2200 798}%
\special{fp}%
%
\special{pn 8}%
\special{pa 2200 996}%
\special{pa 2398 996}%
\special{pa 2398 1194}%
\special{pa 2200 1194}%
\special{pa 2200 996}%
\special{fp}%
%
\special{pn 8}%
\special{pa 2398 996}%
\special{pa 2596 996}%
\special{pa 2596 1194}%
\special{pa 2398 1194}%
\special{pa 2398 996}%
\special{fp}%
%
\special{pn 8}%
\special{pa 2596 996}%
\special{pa 2794 996}%
\special{pa 2794 1194}%
\special{pa 2596 1194}%
\special{pa 2596 996}%
\special{fp}%
%
\special{pn 8}%
\special{pa 2794 996}%
\special{pa 2992 996}%
\special{pa 2992 1194}%
\special{pa 2794 1194}%
\special{pa 2794 996}%
\special{fp}%
%
\special{pn 8}%
\special{sh 0.500}%
\special{pa 2596 1194}%
\special{pa 2794 1194}%
\special{pa 2794 1392}%
\special{pa 2596 1392}%
\special{pa 2596 1194}%
\special{fp}%
%
\special{pn 8}%
\special{pa 2398 1194}%
\special{pa 2596 1194}%
\special{pa 2596 1392}%
\special{pa 2398 1392}%
\special{pa 2398 1194}%
\special{fp}%
%
\special{pn 8}%
\special{pa 2200 1194}%
\special{pa 2398 1194}%
\special{pa 2398 1392}%
\special{pa 2200 1392}%
\special{pa 2200 1194}%
\special{fp}%
%
\special{pn 8}%
\special{pa 2200 1392}%
\special{pa 2398 1392}%
\special{pa 2398 1590}%
\special{pa 2200 1590}%
\special{pa 2200 1392}%
\special{fp}%
%
\special{pn 8}%
\special{pa 2398 1392}%
\special{pa 2596 1392}%
\special{pa 2596 1590}%
\special{pa 2398 1590}%
\special{pa 2398 1392}%
\special{fp}%
%
\special{pn 8}%
\special{pa 2596 1392}%
\special{pa 2794 1392}%
\special{pa 2794 1590}%
\special{pa 2596 1590}%
\special{pa 2596 1392}%
\special{fp}%
%
\special{pn 8}%
\special{pa 2596 1590}%
\special{pa 2794 1590}%
\special{pa 2794 1788}%
\special{pa 2596 1788}%
\special{pa 2596 1590}%
\special{fp}%
%
\special{pn 8}%
\special{pa 2398 1590}%
\special{pa 2596 1590}%
\special{pa 2596 1788}%
\special{pa 2398 1788}%
\special{pa 2398 1590}%
\special{fp}%
%
\special{pn 8}%
\special{pa 2200 1590}%
\special{pa 2398 1590}%
\special{pa 2398 1788}%
\special{pa 2200 1788}%
\special{pa 2200 1590}%
\special{fp}%
%
\special{pn 8}%
\special{pa 2200 1788}%
\special{pa 2398 1788}%
\special{pa 2398 1986}%
\special{pa 2200 1986}%
\special{pa 2200 1788}%
\special{fp}%
%
\special{pn 8}%
\special{pa 2200 1986}%
\special{pa 2398 1986}%
\special{pa 2398 2184}%
\special{pa 2200 2184}%
\special{pa 2200 1986}%
\special{fp}%
%
\special{pn 8}%
\special{pa 2794 1194}%
\special{pa 2992 1194}%
\special{pa 2992 1392}%
\special{pa 2794 1392}%
\special{pa 2794 1194}%
\special{fp}%
%
\special{pn 8}%
\special{pa 2794 1392}%
\special{pa 2992 1392}%
\special{pa 2992 1590}%
\special{pa 2794 1590}%
\special{pa 2794 1392}%
\special{fp}%
%
\special{pn 8}%
\special{pa 2860 1326}%
\special{pa 2992 1326}%
\special{fp}%
\special{sh 1}%
\special{pa 2992 1326}%
\special{pa 2926 1306}%
\special{pa 2940 1326}%
\special{pa 2926 1346}%
\special{pa 2992 1326}%
\special{fp}%
\special{pa 2860 1326}%
\special{pa 2794 1326}%
\special{fp}%
\special{sh 1}%
\special{pa 2794 1326}%
\special{pa 2862 1346}%
\special{pa 2848 1326}%
\special{pa 2862 1306}%
\special{pa 2794 1326}%
\special{fp}%
\special{pa 2662 996}%
\special{pa 2662 1194}%
\special{fp}%
\special{sh 1}%
\special{pa 2662 1194}%
\special{pa 2682 1128}%
\special{pa 2662 1142}%
\special{pa 2642 1128}%
\special{pa 2662 1194}%
\special{fp}%
\special{pa 2662 996}%
\special{pa 2662 600}%
\special{fp}%
\special{sh 1}%
\special{pa 2662 600}%
\special{pa 2642 668}%
\special{pa 2662 654}%
\special{pa 2682 668}%
\special{pa 2662 600}%
\special{fp}%
\special{pa 2464 1326}%
\special{pa 2596 1326}%
\special{fp}%
\special{sh 1}%
\special{pa 2596 1326}%
\special{pa 2530 1306}%
\special{pa 2544 1326}%
\special{pa 2530 1346}%
\special{pa 2596 1326}%
\special{fp}%
\special{pa 2464 1326}%
\special{pa 2200 1326}%
\special{fp}%
\special{sh 1}%
\special{pa 2200 1326}%
\special{pa 2268 1346}%
\special{pa 2254 1326}%
\special{pa 2268 1306}%
\special{pa 2200 1326}%
\special{fp}%
\special{pa 2662 1524}%
\special{pa 2662 1392}%
\special{fp}%
\special{sh 1}%
\special{pa 2662 1392}%
\special{pa 2642 1460}%
\special{pa 2662 1446}%
\special{pa 2682 1460}%
\special{pa 2662 1392}%
\special{fp}%
\special{pa 2662 1524}%
\special{pa 2662 1788}%
\special{fp}%
\special{sh 1}%
\special{pa 2662 1788}%
\special{pa 2682 1722}%
\special{pa 2662 1736}%
\special{pa 2642 1722}%
\special{pa 2662 1788}%
\special{fp}%
\put(30.3000,-13.9000){\makebox(0,0)[lb]{$a(s)$}}%
\put(22.6000,-13.2000){\makebox(0,0)[lb]{$a'(s)$}}%
\put(27.0000,-9.8000){\makebox(0,0)[lb]{$l'(s)$}}%
\put(26.1000,-19.7000){\makebox(0,0)[lb]{$l(s)$}}%
\put(26.6000,-13.4000){\makebox(0,0)[lb]{$s$}}%
\put(25.1000,-24.6000){\makebox(0,0)[lb]{$D(\mu)$}}%
\end{picture}%
 \end{center}
 \caption{Partition and Young diagram.}
 \label{part}
\end{figure}
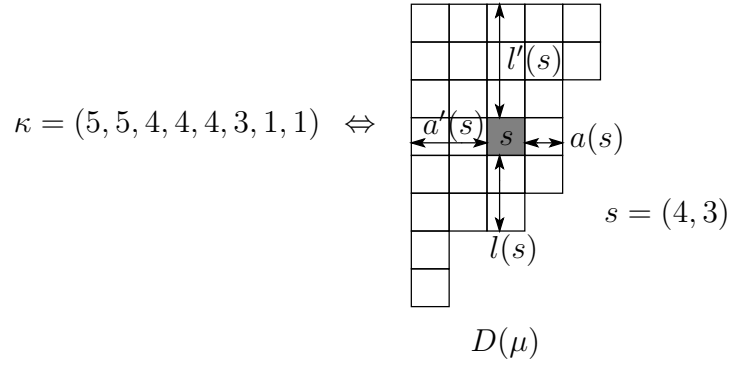

The matrix elements of correlation functions in the Sutherland model are expressed in terms of partitions. 
Partitions can be expressed graphically by  Young diagrams,
whose correspondence with partition is drawn in Figure \ref{part}.
From the top, $\kappa_1$ squares are placed in the first row,
 and $\kappa_2$ squares in the second row, and so on.
Each square is specified by two-dimensional coordinates,
labeling the square at the upper left by $s=(1,1)$.
The first coordinate indicates the vertical position and 
the second the horizontal position.
The length of a partition is defined by the number of nonzero elements
in the partition,
and equal to the length of the first column of the Young diagram.

Some relations of symmetric polynomials 
used in calculating correlation functions are
described in terms of the variables defined on the Young diagram,
$a(s),a'(s),l(s),l'(s)$ for $s=(i,j)$, as
\begin{eqnarray}
 a(s)=\kappa_i-j, \quad &l(s)=\kappa_j'-i, \nonumber\\
 a'(s)=j-1, \quad &\l'(s)=i-1, \nonumber
\end{eqnarray}
where $\kappa_i$ is the $i$-th element of the partition $\kappa$, and
$\kappa'_j$ is the length of $j$-th column 
in the Young diagram (Figure \ref{part}).

In calculating the numerator in the right-hand side of (\ref{hole}),
the term 
$\prod_{i=1}^{N-1}\, (z_i-1)^{\lambda+1}(z_i+1)^{\lambda}$
appears in 
$\tilde{\psi}(0,0)P_{\mbox{g}}^{(2\lambda+1)}(z)$
and it can be expanded by gl$_2$-Jack polynomials. 
The expansion formula is obtained from the 
corresponding relation of Macdonald polynomials\cite{Mac}
\begin{equation}
 \prod_{i=1}^N\, \frac{(z_i;q)}{(tz_i;q)}=\sum_{\mu\in\Lambda_N}\,
 \frac{t^{|\mu|}(t^{-1})_{\mu}^{(q,t)}}{h_{\mu'}(t,q)}\,P_{\mu}(z:q,t),
 \label{expmac}\\
\end{equation}
where
\begin{equation}
 \left(
 \begin{array}{c}
  \displaystyle (r)_{\mu}^{(q,t)}=\prod_{s\in D(\mu)}(t^{l'(s)}-q^{a'(s)}r)\\
  \displaystyle h_{\mu'}(t,q)=\prod_{s\in D(\mu)}(1-q^{a(s)+1}t^{l(s)})
 \end{array}
 \right. .\nonumber
\end{equation}
Taking the limit $q=-p\,,t=-p^{2\lambda+1}\,,p\to1$ for (\ref{expmac}), 
we obtain
\begin{eqnarray}
 \prod_{i=1}^{N-1}\, (1-z_i)^{\lambda+1}(1+z_i)^{\lambda}
 &=
 \!\!\!\!
 \sum_{
  \mu\in\Lambda_{N-1}\atop
  {\scriptsize{\rm s.t.}\,|C_2(\mu)|+|H_2(\mu)|=|\mu|}}
 \!\!\!\!
 P_{\mu}^{(2\lambda+1)}(z) \cdot(-1)^{|\mu|+\sum l'(s)}\,\,
 \nonumber\\
 &\quad\times
 \frac{\displaystyle \prod_{s\in D(\mu)\setminus
 C_2(\mu)}(a'(s)-(2\lambda+1)(l'(s)+1))}
 {\displaystyle \prod_{s\in H_2(\mu)}(a(s)+1+(2\lambda+1)l(s))} \label{exp1},
\end{eqnarray}
where $C_2(\mu),H_2(\mu)$ are the subsets of the squares in Young diagram
$D(\mu)$ defined respectively as (Figure \ref{CH})
\begin{eqnarray}
 &C_2(\mu)=\{s\in D(\mu)|\,a'(s)+l'(s)\equiv 0\,\,\,\,\mbox{mod}\,2\} \\
 &H_2(\mu)=\{s\in D(\mu)|\,a(s)+l(s)+1\equiv 0\,\,\,\, \mbox{mod}\,2\}.
\end{eqnarray}
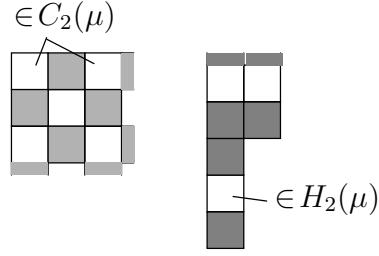
\begin{figure}[t]
 \begin{center}
\unitlength 0.1in
\begin{picture}( 14.2000, 13.0000)(  4.0000,-13.7000)
%
\special{pn 8}%
\special{pa 400 346}%
\special{pa 592 346}%
\special{pa 592 538}%
\special{pa 400 538}%
\special{pa 400 346}%
\special{fp}%
%
\special{pn 8}%
\special{sh 0.300}%
\special{pa 592 346}%
\special{pa 784 346}%
\special{pa 784 538}%
\special{pa 592 538}%
\special{pa 592 346}%
\special{fp}%
%
\special{pn 8}%
\special{pa 784 346}%
\special{pa 976 346}%
\special{pa 976 538}%
\special{pa 784 538}%
\special{pa 784 346}%
\special{fp}%
%
\special{pn 8}%
\special{sh 0.300}%
\special{pa 784 538}%
\special{pa 976 538}%
\special{pa 976 730}%
\special{pa 784 730}%
\special{pa 784 538}%
\special{fp}%
%
\special{pn 8}%
\special{pa 592 538}%
\special{pa 784 538}%
\special{pa 784 730}%
\special{pa 592 730}%
\special{pa 592 538}%
\special{fp}%
%
\special{pn 8}%
\special{sh 0.300}%
\special{pa 400 538}%
\special{pa 592 538}%
\special{pa 592 730}%
\special{pa 400 730}%
\special{pa 400 538}%
\special{fp}%
%
\special{pn 8}%
\special{pa 400 730}%
\special{pa 592 730}%
\special{pa 592 922}%
\special{pa 400 922}%
\special{pa 400 730}%
\special{fp}%
%
\special{pn 8}%
\special{sh 0.300}%
\special{pa 592 730}%
\special{pa 784 730}%
\special{pa 784 922}%
\special{pa 592 922}%
\special{pa 592 730}%
\special{fp}%
%
\special{pn 8}%
\special{pa 784 730}%
\special{pa 976 730}%
\special{pa 976 922}%
\special{pa 784 922}%
\special{pa 784 730}%
\special{fp}%
%
\special{pn 8}%
\special{pa 976 346}%
\special{pa 1040 346}%
\special{fp}%
\special{pa 1040 538}%
\special{pa 976 538}%
\special{fp}%
\special{pa 976 730}%
\special{pa 1040 730}%
\special{fp}%
\special{pa 1040 922}%
\special{pa 976 922}%
\special{fp}%
\special{pa 976 922}%
\special{pa 976 986}%
\special{fp}%
\special{pa 784 986}%
\special{pa 784 922}%
\special{fp}%
\special{pa 592 922}%
\special{pa 592 986}%
\special{fp}%
\special{pa 400 986}%
\special{pa 400 922}%
\special{fp}%
%
\special{pn 8}%
\special{sh 0.300}%
\special{pa 400 922}%
\special{pa 592 922}%
\special{pa 592 986}%
\special{pa 400 986}%
\special{pa 400 922}%
\special{ip}%
%
\special{pn 8}%
\special{sh 0.300}%
\special{pa 976 346}%
\special{pa 1040 346}%
\special{pa 1040 538}%
\special{pa 976 538}%
\special{pa 976 346}%
\special{ip}%
%
\special{pn 8}%
\special{sh 0.300}%
\special{pa 980 730}%
\special{pa 1044 730}%
\special{pa 1044 922}%
\special{pa 980 922}%
\special{pa 980 730}%
\special{ip}%
%
\special{pn 8}%
\special{sh 0.300}%
\special{pa 784 922}%
\special{pa 976 922}%
\special{pa 976 986}%
\special{pa 784 986}%
\special{pa 784 922}%
\special{ip}%
%
\special{pn 8}%
\special{sh 0.500}%
\special{pa 1424 1370}%
\special{pa 1616 1370}%
\special{pa 1616 1178}%
\special{pa 1424 1178}%
\special{pa 1424 1370}%
\special{fp}%
%
\special{pn 8}%
\special{pa 1424 1178}%
\special{pa 1616 1178}%
\special{pa 1616 986}%
\special{pa 1424 986}%
\special{pa 1424 1178}%
\special{fp}%
%
\special{pn 8}%
\special{sh 0.500}%
\special{pa 1424 986}%
\special{pa 1616 986}%
\special{pa 1616 794}%
\special{pa 1424 794}%
\special{pa 1424 986}%
\special{fp}%
%
\special{pn 8}%
\special{sh 0.500}%
\special{pa 1424 794}%
\special{pa 1616 794}%
\special{pa 1616 602}%
\special{pa 1424 602}%
\special{pa 1424 794}%
\special{fp}%
%
\special{pn 8}%
\special{sh 0.500}%
\special{pa 1616 794}%
\special{pa 1808 794}%
\special{pa 1808 602}%
\special{pa 1616 602}%
\special{pa 1616 794}%
\special{fp}%
%
\special{pn 8}%
\special{pa 1616 602}%
\special{pa 1808 602}%
\special{pa 1808 410}%
\special{pa 1616 410}%
\special{pa 1616 602}%
\special{fp}%
%
\special{pn 8}%
\special{pa 1424 602}%
\special{pa 1616 602}%
\special{pa 1616 410}%
\special{pa 1424 410}%
\special{pa 1424 602}%
\special{fp}%
%
\special{pn 8}%
\special{pa 1424 410}%
\special{pa 1424 346}%
\special{fp}%
\special{pa 1616 346}%
\special{pa 1616 410}%
\special{fp}%
\special{pa 1808 410}%
\special{pa 1808 346}%
\special{fp}%
%
\special{pn 8}%
\special{sh 0.500}%
\special{pa 1820 410}%
\special{pa 1628 410}%
\special{pa 1628 346}%
\special{pa 1820 346}%
\special{pa 1820 410}%
\special{ip}%
%
\special{pn 8}%
\special{sh 0.500}%
\special{pa 1616 410}%
\special{pa 1424 410}%
\special{pa 1424 346}%
\special{pa 1616 346}%
\special{pa 1616 410}%
\special{ip}%
\put(4.1000,-2.4000){\makebox(0,0)[lb]{$\in\! C_2(\mu)$}}%
\put(17.7600,-11.9400){\makebox(0,0)[lb]{$\in\! H_2(\mu)$}}%
%
\special{pn 8}%
\special{pa 584 266}%
\special{pa 528 410}%
\special{fp}%
\special{pa 584 274}%
\special{pa 832 402}%
\special{fp}%
\special{pa 1736 1122}%
\special{pa 1560 1098}%
\special{fp}%
\end{picture}%
 \end{center}
 \caption{$C_2(\mu)$ and $H_2(\mu)$. The white squares belong to 
 $C_2(\mu)$ and $H_2(\mu)$ respectively.}
 \label{CH}
\end{figure}
The set $A\setminus B$ means the complementary set of $B$ in $A$.
In the right-hand side of (\ref{exp1}),
the sum of $\mu$ is restricted to the partition satisfying 
 $|C_2(\mu)|+|H_2(\mu)|=|\mu|$.
The numerator of (\ref{hole}) is 
written in terms of the variables of Young diagram and the norm of 
intermediate states as
\begin{eqnarray}
 N&\,
 \!\!\!\!\!\!\!\!
 \sum_{
  \mu\in\Lambda_{N-1}\atop
  {\scriptsize \mbox{s.t.}\,|C_2(\mu)|+|H_2(\mu)|=|\mu|}}
 \!\!\!\!\!\!\!\!
 \delta_{\nu-1/2,\mu-(N/2+\lambda-2)-1/2} \label{eq:mu-nu}\\
 &\times \left(
 \frac{\displaystyle \prod_{s\in D(\mu)/C_2(\mu)}(a'(s)-(2\lambda+1)(l'(s)+1))}
 {\displaystyle \prod_{s\in H_2(\mu)}(a(s)+1+(2\lambda+1)l(s))}\right)^2
 \nonumber\\
 &\times
 \left|\left\{P_{\mu}^{(2\lambda+1)}, P_{\mu}^{(2\lambda+1)}
 \right\}_{N-1,\lambda}\right|^2. 
\end{eqnarray} 
The norm of gl$_2$-Jack polynomial is also obtained by
that of Macdonald polynomial\cite{Mac}
\begin{eqnarray}
 &\left\{P_{\mu}^{(2\lambda+1)}, P_{\mu}^{(2\lambda+1)}
 \right\}_{N,\lambda}=
 \nonumber\\
 &\qquad c_N^{(2\lambda+1,2)}
 \prod_{s\in C_2(\mu)}\frac{a'(s)+(2\lambda+1)(N-l'(s))}
 {a'(s)+1+(2\lambda+1)(N-l'(s)-1)} \nonumber\\
 & \qquad\times
 \prod_{s\in H_2(\mu)}\frac{a(s)+1+(2\lambda+1)l(s)}
 {a(s)+(2\lambda+1)(l(s)+1)},
\end{eqnarray}
where $c_N^{(2\lambda+1,2)}$ is the norm of 
the ground state of the system of $N$ particles 
with spin degrees of freedom $2$
\begin{eqnarray}
 &
 c_N^{(2\lambda+1,2)}=\prod_{1\le i<j\le N}C^{(2\lambda+1)}(j-i),
 \\
 &
 C^{(2\lambda+1)}(k)=\left\{
 \begin{array}{lll}
  &\!\!\!\!\!\!\!\!\!\!
  \frac{\Gamma\left(\frac{(2\lambda+1)(k+1)}
  {2}\right)\Gamma\left(\frac{(2\lambda+1)(k-1)}{2}+1\right)}
  { \Gamma\left(\frac{(2\lambda+1)k}{2}+\frac{1}{2}\right)
  ^{\lower2pt\hbox{\footnotesize{2}}}
  }&k=1 \,\,\mbox{mod}\,2 \\
  \\
  &\!\!\!\!\!\!\!\!\!\!
  \frac{ \Gamma\left(\frac{(2\lambda+1)(k+1)}
  {2}+\frac{1}{2}\right)\Gamma\left(\frac{(2\lambda+1)(k-1)}{2}
  +\frac{1}{2}\right)}
  { \Gamma\left(\frac{(2\lambda+1)k}{2}\right)
  \Gamma\left(\frac{(2\lambda+1)k}{2}+1\right)}
  &k=0 \,\,\mbox{mod}\,2. \\
 \end{array}
 \right.
\end{eqnarray}
To summarize this section, hole propagator is 
written in terms of the variables defined on Young diagram as
\begin{eqnarray}
 G^-(x,t)=&
 \frac{\Gamma((\lambda+1/2)N-\lambda)\Gamma(\lambda+1)}{\Gamma((\lambda+1/2)N)L}
 \!\!\!\nonumber\\  
&\times
 \sum_{
  \mu\in\Lambda_{N-1},{\rm s.t.} S^z_{\rm tot}=+1/2\atop
  |C_2(\mu)|+|H_2(\mu)|=|\mu|}
 \!\!\!e^{-{\rm i}\tilde{\omega}_{\mu}t+{\rm i}\tilde{P}_{\mu}x}\frac{X_\mu^2 Y_\mu(0)}{Y_\mu(\alpha-1/2)Z_\mu(\alpha)Z_\mu(1/2)} 
\label{hole2}
\end{eqnarray}
with $\alpha=1/(2(2\lambda+1))$. $\tilde{\omega}_{\mu}$ and $\tilde{P}_{\mu}$ are, respectively, given by $\tilde{\omega}_{\mu}=\omega_\kappa$ and $\tilde{P}_\mu=P_\kappa$ in (\ref{eq:omegakappa}) through the relation $\mu_i=\alpha_{N-i}-2\kappa_{N-i}-N/2+\lambda-1-i$. The prefactor in front of the summation comes from $c^{(2\lambda+1,2)}_{N-1}/(c^{(2\lambda+1,2)}_{N}L)$. 
We have introduced the following notations:
\begin{eqnarray}
&X_\mu \equiv \prod_{s\in D(\mu)\setminus C_2(\mu)}
\left(-\alpha a'(s)+(l'(s)+1)/2\right),\label{eq:X-def}\\
&Y_\mu(r) \equiv \prod_{s\in C_2(\mu)}
\left(\alpha a'(s)+r +(N-1-l'(s))/2\right),\label{eq:Y-def}\\
&Z_\mu(r) \equiv \prod_{s\in H_2(\mu)}\left(\alpha a(s)+r+l(s)/2\right).\label{eq:Z-def}
\end{eqnarray}

\subsection{Restrictions on $\mu$}
\begin{figure}
\begin{center}
\unitlength 0.1in
\begin{picture}( 10.0000, 16.0000)(  4.0000,-18.1000)
%
\special{pn 8}%
\special{pa 400 210}%
\special{pa 600 210}%
\special{pa 600 410}%
\special{pa 400 410}%
\special{pa 400 210}%
\special{fp}%
%
\special{pn 8}%
\special{sh 0.600}%
\special{pa 400 410}%
\special{pa 600 410}%
\special{pa 600 610}%
\special{pa 400 610}%
\special{pa 400 410}%
\special{fp}%
%
\special{pn 8}%
\special{sh 0.600}%
\special{pa 400 410}%
\special{pa 600 410}%
\special{pa 600 610}%
\special{pa 400 610}%
\special{pa 400 410}%
\special{fp}%
%
\special{pn 8}%
\special{pa 400 610}%
\special{pa 600 610}%
\special{pa 600 810}%
\special{pa 400 810}%
\special{pa 400 610}%
\special{fp}%
%
\special{pn 8}%
\special{sh 0.600}%
\special{pa 600 210}%
\special{pa 800 210}%
\special{pa 800 410}%
\special{pa 600 410}%
\special{pa 600 210}%
\special{fp}%
%
\special{pn 8}%
\special{pa 600 410}%
\special{pa 800 410}%
\special{pa 800 610}%
\special{pa 600 610}%
\special{pa 600 410}%
\special{fp}%
%
\special{pn 8}%
\special{pa 600 410}%
\special{pa 800 410}%
\special{pa 800 610}%
\special{pa 600 610}%
\special{pa 600 410}%
\special{fp}%
%
\special{pn 8}%
\special{pa 600 610}%
\special{pa 800 610}%
\special{pa 800 810}%
\special{pa 600 810}%
\special{pa 600 610}%
\special{fp}%
%
\special{pn 8}%
\special{pa 800 210}%
\special{pa 1000 210}%
\special{pa 1000 410}%
\special{pa 800 410}%
\special{pa 800 210}%
\special{fp}%
%
\special{pn 8}%
\special{sh 0.600}%
\special{pa 800 410}%
\special{pa 1000 410}%
\special{pa 1000 610}%
\special{pa 800 610}%
\special{pa 800 410}%
\special{fp}%
%
\special{pn 8}%
\special{sh 0.600}%
\special{pa 800 410}%
\special{pa 1000 410}%
\special{pa 1000 610}%
\special{pa 800 610}%
\special{pa 800 410}%
\special{fp}%
%
\special{pn 8}%
\special{pa 800 610}%
\special{pa 1000 610}%
\special{pa 1000 810}%
\special{pa 800 810}%
\special{pa 800 610}%
\special{fp}%
%
\special{pn 8}%
\special{pa 400 810}%
\special{pa 600 810}%
\special{pa 600 1010}%
\special{pa 400 1010}%
\special{pa 400 810}%
\special{fp}%
%
\special{pn 8}%
\special{pa 400 1010}%
\special{pa 600 1010}%
\special{pa 600 1210}%
\special{pa 400 1210}%
\special{pa 400 1010}%
\special{fp}%
%
\special{pn 8}%
\special{pa 400 1210}%
\special{pa 600 1210}%
\special{pa 600 1410}%
\special{pa 400 1410}%
\special{pa 400 1210}%
\special{fp}%
%
\special{pn 8}%
\special{pa 400 1410}%
\special{pa 600 1410}%
\special{pa 600 1610}%
\special{pa 400 1610}%
\special{pa 400 1410}%
\special{fp}%
%
\special{pn 8}%
\special{pa 600 810}%
\special{pa 800 810}%
\special{pa 800 1010}%
\special{pa 600 1010}%
\special{pa 600 810}%
\special{fp}%
%
\special{pn 8}%
\special{pa 600 1010}%
\special{pa 800 1010}%
\special{pa 800 1210}%
\special{pa 600 1210}%
\special{pa 600 1010}%
\special{fp}%
%
\special{pn 8}%
\special{pa 600 1210}%
\special{pa 800 1210}%
\special{pa 800 1410}%
\special{pa 600 1410}%
\special{pa 600 1210}%
\special{fp}%
%
\special{pn 8}%
\special{pa 800 810}%
\special{pa 1000 810}%
\special{pa 1000 1010}%
\special{pa 800 1010}%
\special{pa 800 810}%
\special{fp}%
%
\special{pn 8}%
\special{pa 400 1610}%
\special{pa 600 1610}%
\special{pa 600 1810}%
\special{pa 400 1810}%
\special{pa 400 1610}%
\special{fp}%
%
\special{pn 8}%
\special{pa 400 1010}%
\special{pa 600 1010}%
\special{pa 600 1210}%
\special{pa 400 1210}%
\special{pa 400 1010}%
\special{fp}%
%
\special{pn 8}%
\special{pa 400 1010}%
\special{pa 600 1010}%
\special{pa 600 1210}%
\special{pa 400 1210}%
\special{pa 400 1010}%
\special{fp}%
%
\special{pn 8}%
\special{sh 0.600}%
\special{pa 400 810}%
\special{pa 600 810}%
\special{pa 600 1010}%
\special{pa 400 1010}%
\special{pa 400 810}%
\special{fp}%
%
\special{pn 8}%
\special{sh 0.600}%
\special{pa 400 810}%
\special{pa 600 810}%
\special{pa 600 1010}%
\special{pa 400 1010}%
\special{pa 400 810}%
\special{fp}%
%
\special{pn 8}%
\special{pa 400 1010}%
\special{pa 600 1010}%
\special{pa 600 1210}%
\special{pa 400 1210}%
\special{pa 400 1010}%
\special{fp}%
%
\special{pn 8}%
\special{sh 0.600}%
\special{pa 400 1210}%
\special{pa 600 1210}%
\special{pa 600 1410}%
\special{pa 400 1410}%
\special{pa 400 1210}%
\special{fp}%
%
\special{pn 8}%
\special{sh 0.600}%
\special{pa 400 1210}%
\special{pa 600 1210}%
\special{pa 600 1410}%
\special{pa 400 1410}%
\special{pa 400 1210}%
\special{fp}%
%
\special{pn 8}%
\special{pa 400 1410}%
\special{pa 600 1410}%
\special{pa 600 1610}%
\special{pa 400 1610}%
\special{pa 400 1410}%
\special{fp}%
%
\special{pn 8}%
\special{sh 0.600}%
\special{pa 400 1610}%
\special{pa 600 1610}%
\special{pa 600 1810}%
\special{pa 400 1810}%
\special{pa 400 1610}%
\special{fp}%
%
\special{pn 8}%
\special{sh 0.600}%
\special{pa 400 1610}%
\special{pa 600 1610}%
\special{pa 600 1810}%
\special{pa 400 1810}%
\special{pa 400 1610}%
\special{fp}%
%
\special{pn 8}%
\special{sh 0.600}%
\special{pa 600 610}%
\special{pa 800 610}%
\special{pa 800 810}%
\special{pa 600 810}%
\special{pa 600 610}%
\special{fp}%
%
\special{pn 8}%
\special{sh 0.600}%
\special{pa 600 1010}%
\special{pa 800 1010}%
\special{pa 800 1210}%
\special{pa 600 1210}%
\special{pa 600 1010}%
\special{fp}%
%
\special{pn 8}%
\special{sh 0.600}%
\special{pa 800 810}%
\special{pa 1000 810}%
\special{pa 1000 1010}%
\special{pa 800 1010}%
\special{pa 800 810}%
\special{fp}%
%
\special{pn 8}%
\special{sh 0.600}%
\special{pa 800 810}%
\special{pa 1000 810}%
\special{pa 1000 1010}%
\special{pa 800 1010}%
\special{pa 800 810}%
\special{fp}%
%
\special{pn 8}%
\special{pa 1200 210}%
\special{pa 1400 210}%
\special{pa 1400 410}%
\special{pa 1200 410}%
\special{pa 1200 210}%
\special{fp}%
%
\special{pn 8}%
\special{sh 0.600}%
\special{pa 1000 210}%
\special{pa 1200 210}%
\special{pa 1200 410}%
\special{pa 1000 410}%
\special{pa 1000 210}%
\special{fp}%
%
\special{pn 8}%
\special{pa 1000 410}%
\special{pa 1200 410}%
\special{pa 1200 610}%
\special{pa 1000 610}%
\special{pa 1000 410}%
\special{fp}%
%
\special{pn 8}%
\special{pa 1000 410}%
\special{pa 1200 410}%
\special{pa 1200 610}%
\special{pa 1000 610}%
\special{pa 1000 410}%
\special{fp}%
%
\special{pn 8}%
\special{pa 1000 610}%
\special{pa 1200 610}%
\special{pa 1200 810}%
\special{pa 1000 810}%
\special{pa 1000 610}%
\special{fp}%
%
\special{pn 8}%
\special{sh 0.600}%
\special{pa 1000 610}%
\special{pa 1200 610}%
\special{pa 1200 810}%
\special{pa 1000 810}%
\special{pa 1000 610}%
\special{fp}%
\end{picture}%
\caption{Young diagram of a quasi-hole state for $\lambda=2$. In this diagram, $\{\mu'_1,\cdots,\mu'_5\}=\{8,6,4,3,1\}$, $P=\{1,3,4\}$ and $Q=\{2,5\}$. White squares belong to $C_2(\mu)$ and shaded one belong to $D(\mu)\setminus C_2(\mu)$. }
\label{fig:qhyounglambda2}
\end{center}
\end{figure}
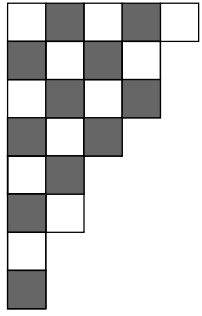

Combining the condition on $\nu$ with that on $\mu$, that is, 
\begin{enumerate}
\item[(i)]$z$ component of total spin of the intermediate state
is larger than that of the ground state added by $1/2$,
\item[(ii)] $|C_2(\mu)|+|H_2(\mu)|=|\mu|$,
\end{enumerate}
there are two conditions for $\mu$ to satisfy in the sum.
On the other hand, 
the factor $a'(s)-(2\lambda+1)(l'(s)+1)$ 
in the product of (\ref{hole2})
implies that the partitions including the square at $(1,2\lambda+2)$
have no contribution in the sum of (\ref{hole2}). 
Hence in the following we consider only the partition with 
$\mu_1\le 2\lambda+1$ and length equal to or shorter than $N-1$.
An example is shown for $\lambda=2$ in Figure~\ref{fig:qhyounglambda2}. 
We define the subsets of $2\lambda+1$ columns of the Young diagram 
by $P$ and $Q$ as follows.
The $j$th column belongs to $P$ when the difference between $j$ and
the length of the $j$th column $\mu_j'$ is odd,
and $Q$ is the complementary set of $P$
\begin{eqnarray}
 P&=\{j\in[1,2\lambda+1]\,|\,\mu_j'-j:\mbox{odd}\} \label{defP} \\
 Q&=\{j\in[1,2\lambda+1]\,|\,\mu_j'-j:\mbox{even}\} \label{defQ}.
\end{eqnarray}
In Figure~\ref{fig:qhyounglambda2}, $\{\mu'_1,\cdots,\mu'_5\}=\{8,6,4,3,1\}$ and hence $P=\{1,3,4\}$ and $Q=\{2,5\}$.
 
We define $n$ as the number of the columns in the set $Q$.
Then $|C_2(\mu)|,|H_2(\mu)|$ are written in terms of $n$ as  
\begin{eqnarray}
 |C_2(\mu)|&=\frac{|\mu|-\lambda+n}{2} \label{cn}
 \\
 |H_2(\mu)|&=\frac{|\mu|-n-\lambda(2\lambda+1)+2(2\lambda\cdot n-n(n-1))}{2}.
 \label{hn}
\end{eqnarray}
These relations are derived in \ref{RE}.
From (\ref{cn}) and (\ref{hn}), the condition $|C_2(\mu)|+|H_2(\mu)|=|\mu|$
can be described as the condition on $n$ by
\begin{eqnarray}
 |C_2(\mu)|+|H_2(\mu)|=|\mu|\,\,\Leftrightarrow\,\,n=\lambda,\lambda+1 
 \label{tspincon}.
\end{eqnarray} 

Next, $z$ component of the total spin for a partition $\mu$
is written as\cite{Ugl}
\begin{eqnarray}
 S_{\mu}&=-|C_2(\mu)|+|D_2(\mu)\setminus C_2(\mu)|+\frac{1}{2} \nonumber\\
 &=\lambda-n+\frac{1}{2}. \label{totalSpin}
\end{eqnarray}
Therefore the condition for total spin
is also written in terms of $n$ by
\begin{equation}
 S_{\mu}=1/2\,\,\Leftrightarrow\,\,
 n=\lambda. 
 \label{spincon}
\end{equation}
The relation (\ref{totalSpin}) 
gives the meaning of the relation (\ref{tspincon})
that intermediate states $\mu$
that arise by acting $(\psi_{\downarrow}+\psi_{\uparrow})$ on 
the wave function of the ground state
contain only the states with $S_{\mu}=+1/2$
or $-1/2$.
Then the relation (\ref{spincon}) imposes the restriction to take only 
one of the annihilation operator.
Putting together all conditions,
the sum of $\mu$ is taken over the partitions satisfying
\begin{eqnarray}
  &\mu_1\le 2\lambda+1\mbox{ and length equal to or shorter than }N-1
   \label{con1}
   \\
  &n=\lambda.
   \label{con2}
\end{eqnarray}

\section{Quasi-hole Description of Hole Propagator}\label{Q}
In this section, we rewrite the expression for hole propagator in terms of rapidities and spins of quasi-holes.

\subsection{Rapidities and spins of quasi-holes}
From (\ref{con1}), $N-1$ particle states relevant to the hole propagator (\ref{hole2}) 
are parameterized by the set of length of each column 
$\left\{\mu'_1,\cdots,\mu'_{2\lambda+1}\right\}$ 
and the set of ``spin variables''
$$\left\{\sigma_1,\cdots,\sigma_{2\lambda+1}\right\},$$
the entry of which is defined by
\begin{equation}
\sigma_j=\left\{
\begin{array}{rc}
1/2,& j\in P\\
-1/2,& j\in Q\\
\end{array}
\right.
\label{eq:sigma-j}
\end{equation}
for $i \in [1,2\lambda+1]$. 
For later convenience, we introduce auxiliary notations $\mu'_0=N-1$, $\mu'_{2\lambda+2}=0$ and  
\begin{equation}
\sigma_0=1/2,\quad \sigma_{2\lambda+2}=-1/2.
\label{eq:sigma0-2lambda+1}
\end{equation}
Regarding the length $\mu'_0$ of ``$0$-th'' column as $N-1$, 
which is odd and that of $2\lambda+2$-th as 0, 
the definition (\ref{eq:sigma0-2lambda+1}) 
is a natural extension of (\ref{eq:sigma-j}). 

Furthermore, we introduce the renormalized momentum
\begin{equation}
\tilde{\mu}'_j=\mu'_j -\frac{N-1}{2}+\frac{\lambda+1-j}{2\lambda+1}
\label{eq:mutildej}
\end{equation}
for $j\in [0,2\lambda+2]$. In terms of (\ref{eq:sigma-j}) and (\ref{eq:mutildej}) for $j\in [1, 2\lambda+1]$, excitation energy of eigenstates relevant to hole propagator is written in the form of free particles. 
Matrix element appearing in hole propagator is written in terms of $\{\tilde{\mu}_j,\sigma_j\}$ for $j\in [0,2\lambda+2]$. 
 We will see in the following that $\tilde{\mu}'_j+\sigma_j$ and $\sigma_j$ can be interpreted as the rapidity and spin of $j$-th quasi-hole, respectively. 

\subsection{Hole propagator in finite-sized systems}
The excitation energy $\tilde{\omega}_{\mu}\equiv \omega_{\kappa}$ and the momentum $\tilde{P}_{\mu}\equiv P_{\kappa}$
are described by  
\begin{eqnarray}
\tilde{\omega}_\mu&=-(2\lambda+1)\sum_{j=1}^{2\lambda+1}
\left(\frac{\pi(\tilde{\mu}'_j+\sigma_j)}{L}\right)^2
+\frac{4\pi^2\lambda(\lambda+1)}{3L^2}\label{eq:tildeomega}\\
\tilde{P}_\mu&=(2\pi/L)\sum_{i=1}^{N-1}\kappa_i=-\sum_{j=1}^{2\lambda+1}(\pi(\tilde{\mu}'_j+\sigma_j)/L). \label{eq:tildeP}
\end{eqnarray}
The derivation of (\ref{eq:tildeomega}) and (\ref{eq:tildeP}) is given in \ref{ME}.
The matrix element in the hole propagator (\ref{hole2})
can be described in terms of the renormalized momenta and spin variables as the energy spectrum.

$X_\mu$, $Y_\mu(r)$ and $Z_\mu(r)$ defined in (\ref{eq:X-def}), (\ref{eq:Y-def}) and (\ref{eq:Z-def}), respectively, are described as
\begin{equation}
X_\mu=\prod_{j=1}^{2\lambda+1}
\frac{\Gamma[(\tilde{\mu}_j'-\tilde{\mu}_{2\lambda+2}'+1
-\delta_{\sigma_{j}\sigma_{2\lambda+2}})/2]}{\Gamma[j/(2\lambda+1)]},
\label{eq:Xmu-final}
\end{equation}
\begin{equation}
Y_\mu(r)=\prod_{j=1}^{2\lambda+1}
\frac{\Gamma[\alpha+r+N/2+(j-1)/(2\lambda+1)]}
{\Gamma[(\tilde{\mu}'_0-\tilde{\mu}'_j +1+\delta_{\sigma_0\sigma_j})
/2-\alpha+r]}, 
\label{eq:Ymu-final}
\end{equation}
\begin{eqnarray}
Z_\mu(r)&=(\Gamma[r+1/2])^{-(2\lambda+1)}\nonumber\\
&\times\prod_{j=1}^{2\lambda+1}
\Gamma[(\tilde{\mu}'_j -\tilde{\mu}'_{2\lambda+2}
-\delta_{\sigma_j\sigma_{2\lambda+2}})/2+r-\alpha+1/2]\nonumber\\
&\times\prod_{1\le j<k\le 2\lambda+1}
\frac{\Gamma[(\tilde{\mu}'_j -
\tilde{\mu}'_k-\delta_{\sigma_j\sigma_k})/2+r-\alpha+1/2]}
{\Gamma[(\tilde{\mu}'_j -\tilde{\mu}'_k+\delta_{\sigma_j\sigma_k})/2+r] }, 
\label{eq:zmu-r}
\end{eqnarray}
as shown in \ref{ME}. 
With use of (\ref{eq:Xmu-final}), (\ref{eq:Ymu-final}) 
and (\ref{eq:zmu-r}), 
the expression (\ref{hole2}) is rewritten as
\begin{eqnarray}
G^-(x,t)=
 K_\lambda(N) d\sum_{
  0\le \mu'_{2\lambda+1}\le\cdots\le\mu'_1\le N-1
      }&
 \sum_{\{\sigma_j\}}
\delta_{(\sum_j \sigma_j),1/2}
\exp[-i\tilde{\omega}_{\mu}t+i\tilde{P}_{\mu}x]\nonumber\\ 
&\times F(\{\tilde{\mu}'_j, \sigma_j\}), \nonumber\\
\label{holefinite-final}
\end{eqnarray}
with the particle density $d=N/L$ and the constant $K_\lambda(N)$
\begin{eqnarray}
K_\lambda(N)&=\frac{\Gamma((\lambda+1/2)N-\lambda)\Gamma(\lambda+1)}{\Gamma((\lambda+1/2)N-\lambda)}\frac{(\Gamma[((\lambda+1)/(2\lambda+1))])^{2\lambda+1}}
{\prod_{j=1}^{2\lambda+1}(\Gamma[j/(2\lambda+1)])^2}\nonumber\\
&\times\prod_{j=1}^{2\lambda+1}
\frac{\Gamma[j/(2\lambda+1)+N/2-\alpha]}
{\Gamma[j/(2\lambda+1)+N/2-1/2]}
\end{eqnarray}
and form factor
\begin{eqnarray}
&F(\{\tilde{\mu}'_j, \sigma_j\})\nonumber\\
&=
\prod_{j=1}^{2\lambda+1}
\frac{\Gamma[(\tilde{\mu}'_0-\tilde{\mu}'_j+\delta_{\sigma_j\sigma_0})/2]}
{\Gamma[(\tilde{\mu}'_0-\tilde{\mu}'_j+1+\delta_{\sigma_j\sigma_0})/2-\alpha]}
\nonumber\\
&\times\prod_{j=1}^{2\lambda+1}
\frac{\Gamma[(\tilde{\mu}'_j-\tilde{\mu}'_{2\lambda+2}+1-
\delta_{\sigma_j\sigma_{2\lambda+2}})/2]}
{\Gamma[(\tilde{\mu}'_j-\tilde{\mu}'_{2\lambda+2}+1-
\delta_{(\sigma_j\sigma_{2\lambda+2}})/2+1/2-\alpha]}\nonumber\\
&\times\prod_{1\le j<k\le 2\lambda+1}
\frac{\Gamma[(\tilde{\mu}'_j-\tilde{\mu}'_k+\delta_{\sigma_j\sigma_k})/2+
\alpha]\Gamma[(\tilde{\mu}'_j-\tilde{\mu}'_k+\delta_{\sigma_j\sigma_k})/2+1/2]}
{\Gamma[(\tilde{\mu}'_j-\tilde{\mu}'_k-
\delta_{\sigma_j,\sigma_k})/2+1/2]
\Gamma[(\tilde{\mu}'_j-\tilde{\mu}'_k-\delta_{\sigma_j\sigma_k})/2+1-\alpha]}.\nonumber\\
\end{eqnarray}

\subsection{Thermodynamic limit}

Changing the variables to new ones by 
\begin{eqnarray}
 \frac{\mu_j'}{N}= \frac{1-u_j}{2},
\end{eqnarray}
which are finite in the thermodynamic limit.
Using the relation $\Gamma(N+a)/\Gamma(N+b)\to N^{a-b}\,(N\to\infty)$,
(\ref{hole2}) converges to an expression with finite value.
The result in the thermodynamic limit 
$N\to\infty(N/L=\mbox{const.})$ is written as
\begin{eqnarray}
 G^{-}(x,t)
 &=K'_\lambda d \left(\prod_{i=1}^{2\lambda+1}\sum_{\sigma_i=\pm 1/2}\int_{-1}^{1}{\rm d}u_i\right)\delta_{(\sum_{i=1}^{2\lambda+1} \sigma_i), 1/2}\nonumber\\
 &\,\,\,\times
 \prod_{j=1}^{2\lambda+1}
 \left(1-u_j^2\right)^{-\lambda/(2\lambda+1)}
\prod_{
j<k
}^{2\lambda+1}
 \left|u_k-u_j\right|^{2\delta_{\sigma_j,\sigma_k}-2\lambda/(2\lambda+1)} \nonumber\\
 &\,\,\,\quad \times\exp\left[
 i\left((2\lambda+1)\left(\frac{\pi d}{2}\right)^2
 \sum_{j=1}^{2\lambda+1}u_j^2\right)t+
 i\left(\frac{\pi d}{2}\sum_{j=1}^{2\lambda+1}u_j\right)x
 \right],\nonumber\\
 \label{holeres}
\end{eqnarray}
with the overall constant
\begin{equation}
K'_\lambda=\frac{\Gamma(\lambda+1)}
 {4\,(2\lambda+1)^{\lambda}\,\Gamma(2\lambda+2)}
 \prod_{j=1}^{2\lambda+1}\,\,
 \frac{\displaystyle\Gamma\left((\lambda+1)/(2\lambda+1)\right)}
 {\displaystyle\Gamma\left(j/(2\lambda+1)\right)^2}.  
\end{equation}
The exponents of interparticle part of the matrix element
in (\ref{holeres})
depend on whether spins of a pair of particles are parallel or anti-parallel. The result coincides with the earlier result\cite{Kat,Ka2}.

\section{Spectral Function for $\lambda=1$\label{S}}
\begin{figure}[t]
 \begin{center}
  \begin{tabular}{cc}
   \begin{minipage}{0.6\hsize}
 \begin{tabular}{ccc}
  \begin{minipage}{0.3\hsize}
   \includegraphics[width=30mm]{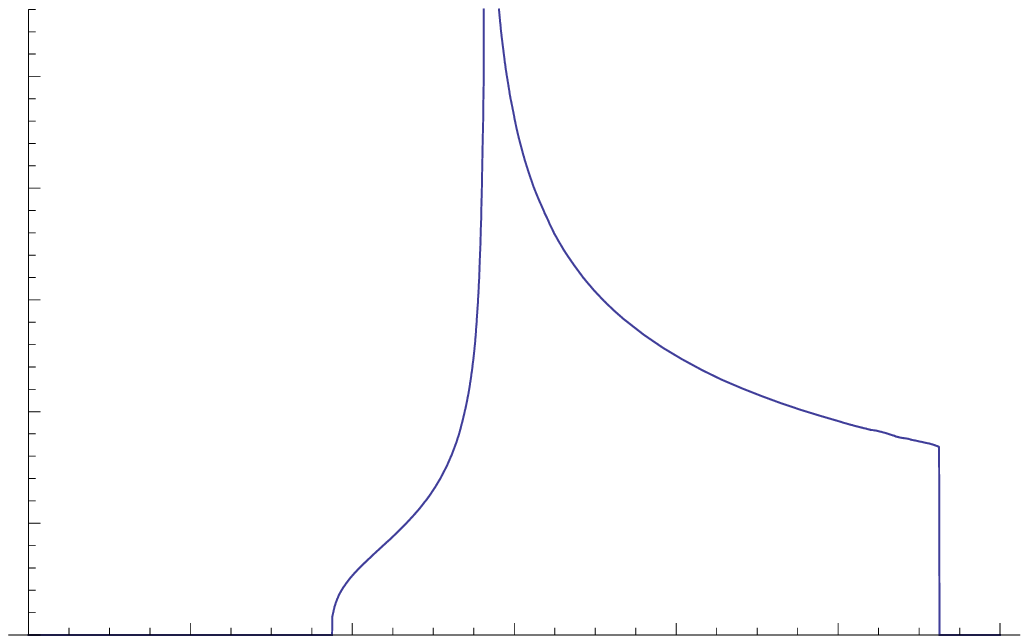}
  \end{minipage}
  \begin{minipage}{0.3\hsize}
   \includegraphics[width=30mm]{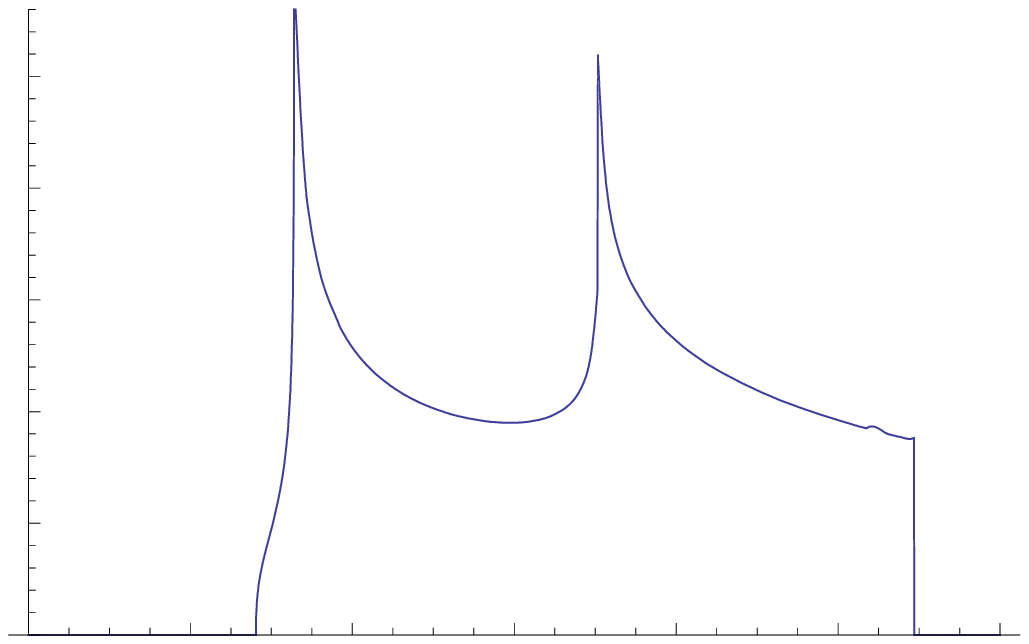}
  \end{minipage}
  \begin{minipage}{0.3\hsize}
   \includegraphics[width=30mm]{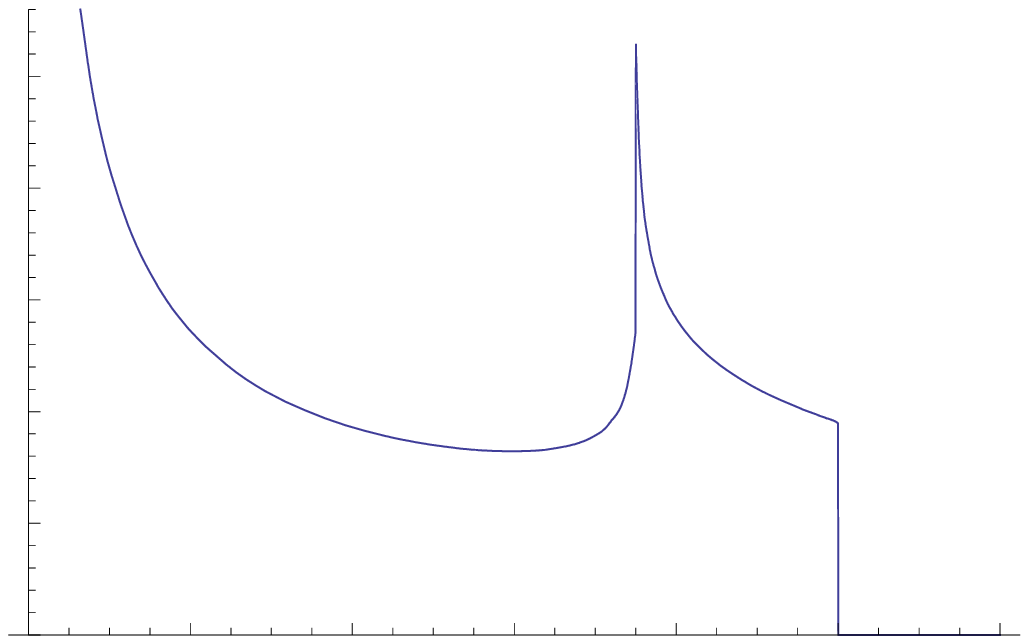}
  \end{minipage}
  \end{tabular}
  \vspace{1mm}
  \begin{tabular}{ccc}
  \begin{minipage}{0.3\hsize}
   \begin{center}
    \footnotesize{\sf{(a)}}
   \end{center}
  \end{minipage}
  \begin{minipage}{0.3\hsize}
   \begin{center}
    \footnotesize{\sf{(b)}}
   \end{center}
  \end{minipage}
  \begin{minipage}{0.3\hsize}
   \begin{center}
    \footnotesize{\sf{(c)}}
   \end{center}
  \end{minipage}
  \end{tabular}
  \vspace{0mm}
  \begin{tabular}{ccc}
  \begin{minipage}{0.3\hsize}
   \includegraphics[width=30mm]{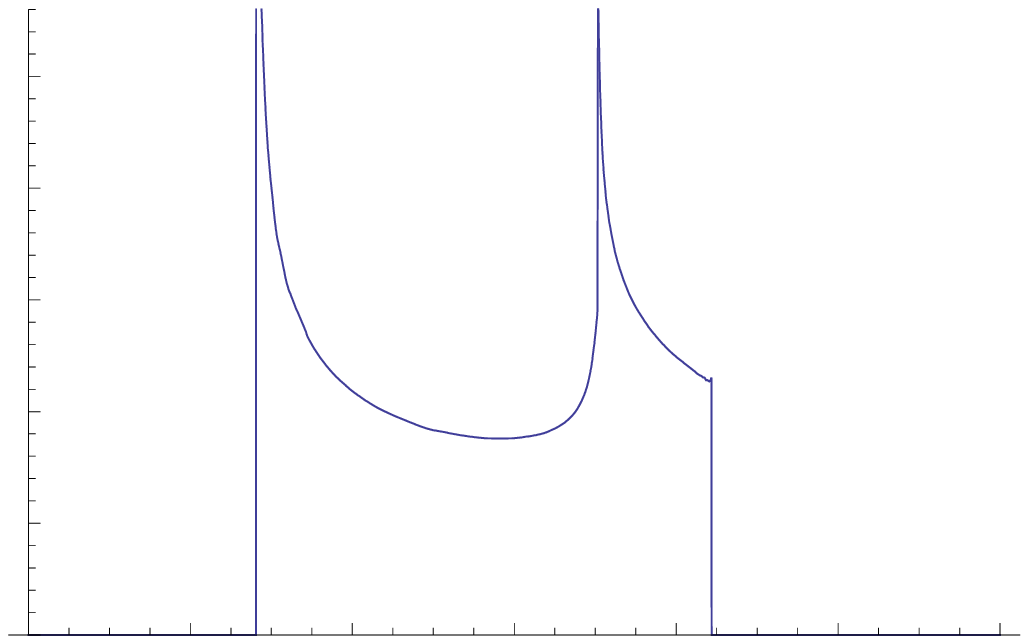}
  \end{minipage}
  \begin{minipage}{0.3\hsize}
   \includegraphics[width=30mm]{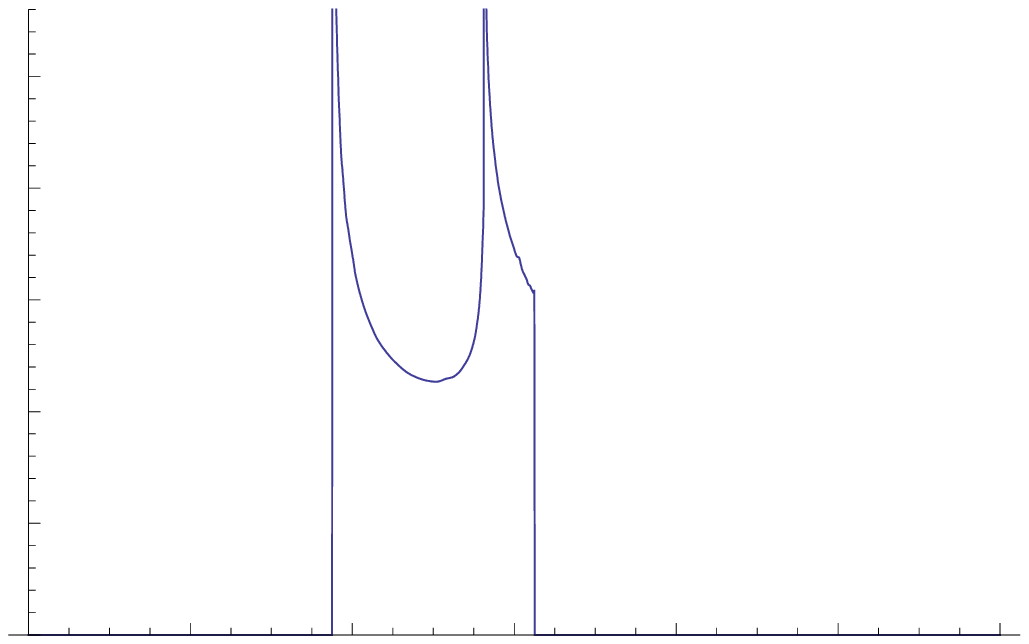}
  \end{minipage}
  \begin{minipage}{0.3\hsize}
   \includegraphics[width=30mm]{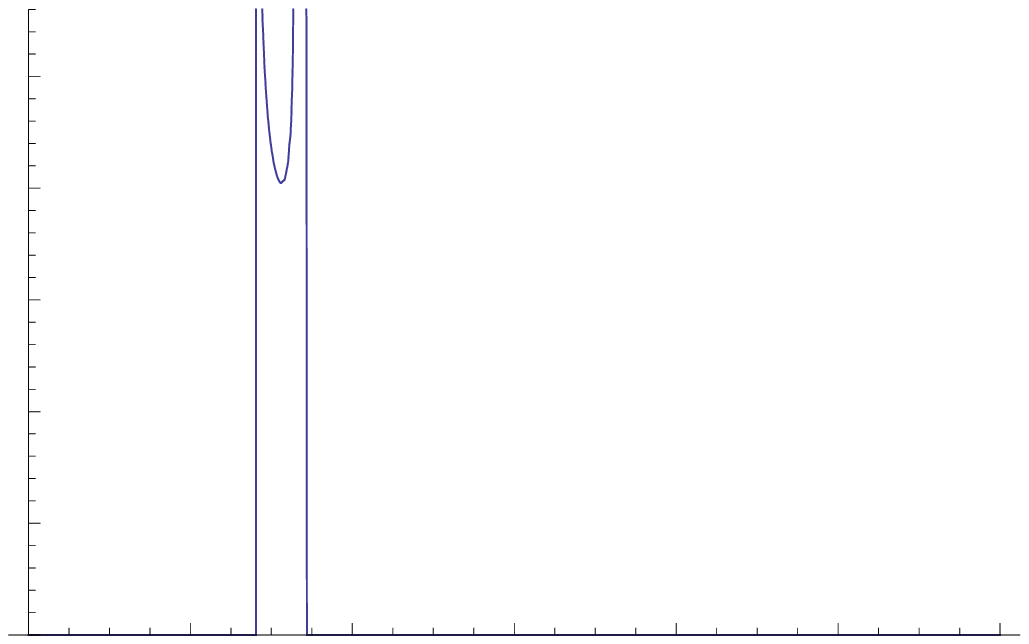}
  \end{minipage}
  \end{tabular}
  \vspace{1mm}
  \begin{tabular}{ccc}
  \begin{minipage}{0.3\hsize}
   \begin{center}
    \footnotesize{\sf{(d)}}
   \end{center}
  \end{minipage}
  \begin{minipage}{0.3\hsize}
   \begin{center}
    \footnotesize{\sf{(e)}}
   \end{center}
  \end{minipage}
  \begin{minipage}{0.3\hsize}
   \begin{center}
    \footnotesize{\sf{(f)}}
   \end{center}
  \end{minipage}
  \end{tabular}
    \end{minipage}
    \begin{minipage}{0.4\hsize}
     \vspace{7.7mm}
   \includegraphics[width=55mm]{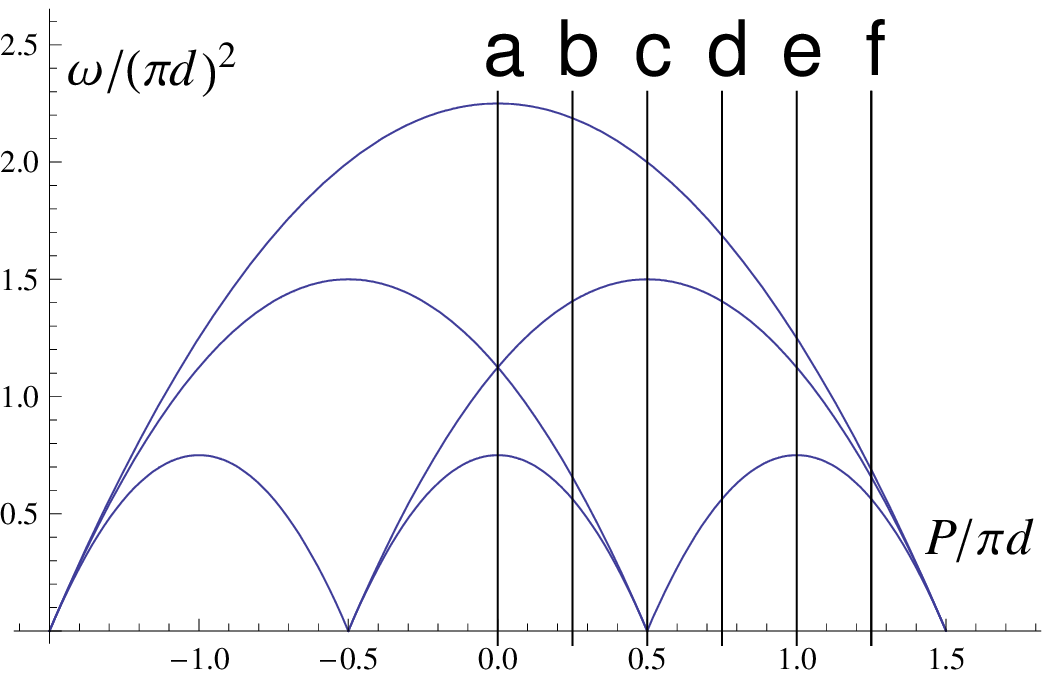}
    \end{minipage}
   \end{tabular}
   \caption{Spectral function of spin $1/2$ 
  Calogero-Sutherland model for $\lambda=1$.}
  \label{sp}
  \end{center}
\end{figure}

Spectral function for hole propagator is defined as a Fourier transformation
of $G^-(x,t)$:
\begin{eqnarray}
 A^{-}(p,\epsilon)=\frac{1}{2\pi}
 \int_{-\infty}^{\infty} {\rm d}x\int_{-\infty}^{\infty}{\rm d}t 
 \,e^{{\rm i}(\epsilon-\mu)t-{\rm i}Px}
 G^-(x,t).
\end{eqnarray}
We calculate the spectral function at $\lambda=1$, 
which is the easiest nontrivial case in this model.
The spectral function has non-zero value at finite area in 
energy-momentum plane (Figure \ref{sp}).
The area is enclosed by four parabolic lines,
and each of them can be interpreted in terms of quasi-hole picture.
The upper edge $\epsilon=-p^2+9(\pi d)^2/4$ with $|p|<3\pi d/2$ corresponds to 
the excited state with three quasi-holes having a same momentum.
The three lower edges 
\begin{eqnarray}
&\epsilon=-3p^2+3(\pi d)^2/4,\quad |p|<\pi d/2,\nonumber\\
&\epsilon=-3(p+\pi d)^2+3(\pi d)^2/4,\quad -3\pi d/2 <p<-\pi d/2\nonumber\\
&\epsilon=-3(p-\pi d)^2+3(\pi d)^2/4,\quad \pi d/2 <p<3\pi d/2\nonumber
\end{eqnarray}
correspond
to the states that one quasi-hole is excited while the other two quasi-holes reside at Fermi points. 

The intensity of the spectral function is also drawn in Figure \ref{sp}.
The spectral function diverges at 
two lower edges $\epsilon=-3(p\pm\pi d)^2+3(\pi d)^2/4$ for 
$\pi d/2 <|p|<3\pi d/2$ and
two other parabolic lines $\epsilon=-3(p\mp\pi d/2)^2/2+3(\pi d)^2/2$, 
with $-\pi d/2 \le p \le 3\pi d/2$ for upper sign 
and $-3\pi d/2\le p\le -\pi d/2$ for lower sign. 
At the lower edge $\epsilon=-3p^2+3(\pi d)^2/4$ 
for $-\pi d/2 \le p \le \pi d/2$,
the spectral function becomes zero and arises as
$(\epsilon+3p^2-3(\pi d)^2/4)^{1/3}$,
and at $\epsilon=-3(p\pm\pi d)^2+3(\pi d)^2/4$ 
it diverges as $(\epsilon+3(p\pm\pi d)^2-3(\pi d)^2/4)^{-1/3}$.
At the middle line, spectral function diverges as
$|\epsilon+3(p\mp\pi d/2)^2/2-3(\pi d)^2/2|^{-1/6}$, which will be derived in \ref{SW}.
At upper edge, the spectral function takes finite value
proportional to $(1-(2p/3\pi d)^2)^{-1}$.
Arikawa pointed out \cite{Ari} that the hole propagator of SU($2$) Calogero-Sutherland model for $\lambda=1$
is equivalent to the dynamical color correlation function of SU($3$) 
Haldane-Shastry model\cite{Yamamoto2000,Yamamoto1999,Ari}.
Although the singularities of the upper edge and lower edges have been obtained in earlier works\cite{Yamamoto1999,Ari},
our calculation shows that the spectral function at the middle line 
diverges with exponent $1/6$, contrary to \cite{Yamamoto1999}.

\section{Conclusion}

To calculate the particle propagator of spin Calogero-Sutherland model
by Uglov's method,
we transformed the field annihilation operator
on Yangian Gelfand-Zetlin basis by the mapping $\Omega$
and
proved that it becomes also the field annihilation operator
on gl$_2$-Jack polynomials.
This ensures the possibility of calculating 
1-particle Green's function using the Uglov's method.

Next, with use of this method,
we calculated hole propagator
for non-negative integer interaction parameter
by taking the restricted product over Young diagrams of intermediate states.
Thermodynamic limit of hole propagator was also taken,
and we confirmed that the result obtained here coincides with
that of the former results with Jack polynomial with prescribed symmetry.
Spectral function for $\lambda=1$ was calculated,
and drawn in energy-momentum plane.
There appear the divergences of intensity
on one and two quasi-hole excitation lines.

The calculation of particle propagator
will be published in a separate paper.

\ack
We are grateful to M. Arikawa for giving us the information about 
the dynamical correlation function of SU($3$) Haldane-Shastry model.
We also thank Y. Nagai 
for his help in drawing the spectral function in Sec. \ref{S}.
This work was supported in part by 
Global COE Program "the Physical Sciences Frontier", MEXT, Japan.

\appendix
\setcounter{section}{0}
\section{Restriction on the Product}\label{RE}

The number of the squares of the subset of the Young diagram, $|C_2(\mu)|$
and $|H_2(\mu)|$, can be uniquely described in terms of $n$, the number of 
the columns which belong to $Q$. 
This can be verified by describing $|C_2(\mu)|-|D(\mu)\setminus C_2(\mu)|$
and $|H_2(\mu)|-|D(\mu)\setminus H_2(\mu)|$ in terms of $n$,
where $A\setminus B$ means the complement of $B$ in $A$.

\begin{figure}[t]
 \begin{center}
  \input{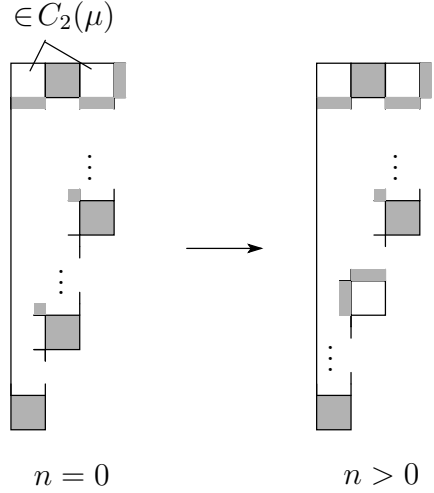}
 \end{center}
 \caption{The relation between the number of the squares of 
 $C_2(\mu)$ and $D(\mu)\setminus C_2(\mu)$ and $n$.
 White squares are the elements of $C_2(\mu)$,
 and shaded squares are that of $D(\mu)\setminus C_2(\mu)$.
 For $n=0$, the bottom of each column are shaded squares
 (the elements of the $D(\mu)\setminus C_2(\mu)$),
 and for $n>0$,
 there appears the white square (the elements of the $C_2(\mu)$) 
 in $n$ columns.} 
 \label{Cn2}
\end{figure}

In each column and each row of the Young diagram,
the element of $C_2(\mu)$ and that of $D(\mu)\setminus C_2(\mu)$
are aligned alternately, 
and at the upper left of the diagram $s=(1,1)\in C_2(\mu)$.
From the definition of the set $P$ and $Q$ ((\ref{defP}) and (\ref{defQ})),
$j$-th column with $j\in P$ has
even number of squares for odd $j$
and odd number for even $j$.
Thus the square at the bottom of the $j$-th column in $P$ is an element of 
$D(\mu)\setminus C_2(\mu)$.
For $n=0$, all the columns belong to the set $P$,
and therefore $|C_2(\mu)|-|D(\mu)\setminus C_2(\mu)|=-\lambda$.
The change of a column in $P$ to $Q$ increases $|C_2(\mu)-D(\mu)/C_2(\mu)|$ by one (Figure \ref{Cn2}). We thus obtain
\begin{eqnarray}
 |C_2(\mu)|-|D(\mu)\setminus C_2(\mu)|=-\lambda+n.
\end{eqnarray}

\begin{figure}[t]
 \begin{center}
  \input{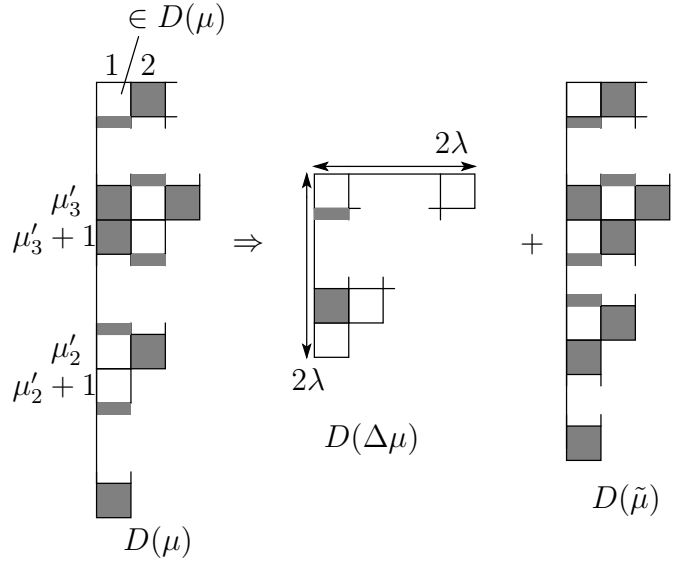}
 \end{center}
 \caption{Young diagram $D(\mu)$ is decomposed to  two diagrams
 $D(\Delta\mu)$ and $D(\tilde{\mu})$. 
 White squares are the elements of $H_2(\mu)$,
 and shaded squares are that of $D(\mu)\setminus H_2(\mu)$.
 In each column of $D(\Delta\mu)$, 
 the element of $H_2(\mu)$ and that of $D(\mu)\setminus H_2(\mu)$
 are aligned alternately.} 
 \label{dec2}
\end{figure}

Next we describe $|H_2(\mu)|-|D(\mu)\setminus H_2(\mu)|$ in terms of $n$.
First we consider a partition having no adjacent columns with the same length,
and then we consider the general case.
In each column, the square at the bottom $s=(\mu_j',j)$ is an element of 
$D(\mu)\setminus H_2(\mu)$.
The element of $H_2(\mu)$ and that of $D(\mu)\setminus H_2(\mu)$
are aligned alternately except for several rows. 
The exceptions occur at the $\mu_k'$-th row and $\mu_k'+1$-th row 
in $j$-th column with $k=[j+1,j+2,\cdots,2\lambda+1]$,
where two elements of the same subset 
$H_2(\mu)$ or $D(\mu)\setminus H_2(\mu)$
are aligned vertically (Figure \ref{dec2}).
Thus we remove the $\mu_j'+1$-th row with $j=[2,3,\cdots,2\lambda+1]$
from the original Young diagram,
and make the diagrams $D(\tilde{\mu})$
in which the element of $H_2(\mu)$ and that of $D(\mu)\setminus H_2(\mu)$
are aligned alternately without any exceptions in each column.
The removed $2\lambda$ rows make a new diagram $D(\Delta\mu)$ 
with length $2\lambda$ and $(\Delta\mu)_1=2\lambda$ (Figure \ref{dec2}).

For $n=0$, 
each column of $D(\tilde{\mu})$ has even number of squares.
Therefore, the contribution to $|H_2(\mu)|-|D(\mu)\setminus H_2(\mu)|$
from $\tilde{\mu}$ is given by
\begin{eqnarray}
 |H_2(\mu)|_{\tilde{\mu}} -|D(\mu)\setminus H_2(\mu)|_{\tilde{\mu}} = 0.
\end{eqnarray}
When a column in $D(\mu)$ changes from $P$ to $Q$, the number of columns in $D(\tilde{\mu})$ with odd length increases by one. 
We thus obtain
\begin{eqnarray}
 |H_2(\mu)|_{\tilde{\mu}} -|D(\mu)\setminus H_2(\mu)|_{\tilde{\mu}} = -n\label{eq:contribution-tildemu}
\end{eqnarray}
for $n\ge 0$.

Now we consider the contribution of $D(\Delta \mu)$ to $|H_2(\mu)| -|D(\mu)\setminus H_2(\mu)|$. 
The square $s=(j,k)$ with $1\le j\le 2\lambda\,,\,$
$1\le k\le 2\lambda+1-j$ in $D(\Delta \mu)$ 
comes from $s'=(\mu'_{2\lambda+2-j}+1,k)$ in $D(\mu)$. 
With use of $a(s')=2\lambda+2-j-k-1$ and 
$l(s')=\mu'_{k}-\mu'_{2\lambda+2-j}-1$, it follows that 
\begin{equation}
a(s')+l(s')+1\equiv 
\left\{
\begin{array}{cc}
1&(k,2\lambda+2-j)\in (P,P)\mbox{ or }(Q,Q)\\
0&(k,2\lambda+2-j)\in (P,Q)\mbox{ or }(Q,P)\\
\end{array}
\right. ,
\end{equation}
and hence $s=(j,k)\in H_2(\mu)$ in $D(\Delta \mu)$ $\Leftrightarrow$
$(k,2\lambda+2-j)\in (P,Q)\mbox{ or }(Q,P)$.
Since two coordinates $k$ and $2\lambda+2-j$ vary with 
$1\le k < 2\lambda+2-j\le 2\lambda+1$,
\begin{eqnarray}
 &|H_2(\mu)|_{\Delta\mu}=n(2\lambda+1-n) \\
 &|D(\mu)\setminus H_2(\mu)|_{\Delta\mu}
 =n(n-1)/2+(2\lambda+1-n)(2\lambda-n)/2
\end{eqnarray}
and 
\begin{eqnarray} 
 |H_2(\mu)|_{\Delta\mu}-|D(\mu)\setminus H_2(\mu)|_{\Delta\mu}
 =-2n^2+2(2\lambda+1)n-\lambda(2\lambda+1)\label{eq:contribution-Deltamu}
\end{eqnarray}
holds.
From (\ref{eq:contribution-tildemu}) and (\ref{eq:contribution-Deltamu}), we obtain
\begin{eqnarray}
 |H_2(\mu)|-|D(\mu)\setminus H_2(\mu)|= 
 -2n^2+(4\lambda+1)n-\lambda(2\lambda+1). \label{appH}
\end{eqnarray}
Next, we apply the result to 
a partition having two or more columns with the same length.
When a partition have $m$ columns with same length,
we extract the $2[m/2]$ neighboring columns from the original partition,
where $[n]$ means the maximum integer that does not exceed $n$.
Since the extracted $2[m/2]$ columns have the same number of the elements of 
$H_2(\mu)$ and $D(\mu)\setminus H_2(\mu)$,
there is no contribution to $|H_2(\mu)|-|D(\mu)\setminus H_2(\mu)|$.
We can use the result (\ref{appH}) for the 
left partition with $2\lambda+1-2[m/2]$ columns and obtain
the equation by replacing $\lambda$ in (\ref{appH}) by $\lambda-[m/2]$.
The above discussion can be applied to a partition that has 
more than one set of columns with the same length.
These applications, however, do not change the result on the condition
$|C_2(\mu)|+|H_2(\mu)|=|\mu|$.

\section{Energy Spectrum and Matrix Elements}\label{ME}
In this appendix, we derive (\ref{eq:tildeomega}) and (\ref{eq:tildeP}) in \ref{Es-ME} and (\ref{eq:Xmu-final}), (\ref{eq:Xmu-final}) and (\ref{eq:zmu-r}) in \ref{Me-ME}.

\subsection{Energy spectrum}\label{Es-ME}
In the excitation energy $\omega_\kappa$ in (\ref{eq:omegakappa}), 
the energy  
$$
E_{N-1}(\kappa)=(2\pi/L)^2\sum_{i=1}^{N-1}\left(\kappa_i +\lambda(N-2i)/2\right)^2$$ 
of $N-1$ particle state $\kappa$ is rewritten as
\begin{equation}
E_{N-1}(\kappa)=(\pi/L)^2\sum_{i=1}^{N-1}\left(-\mu_i+\lambda+(1+2\lambda)\left(i-N/2\right)+\alpha_{N-i}-1\right)^2,\label{eq:EN-1}
\end{equation}
with use of 
\begin{equation}
\nu_i=\alpha_{N-i}-2\kappa_{N-i}-N+1+i
\label{nu-i-kappa}
\end{equation}
(which is (\ref{parttrans}) with replacement of $N$ by $N-1$)
and 
\begin{equation}
\mu=\nu+N/2+\lambda-2
\end{equation}
coming from (\ref{eq:mu-nu}). 
In (\ref{eq:EN-1}), the value of $\alpha_{N-i}$ is given by
\begin{equation}
\alpha_{N-i}=\left\{
\begin{array}{cc}
2,& (i,\mu_i)\in C_2(\mu)\\
1,& (i,\mu_i)\in D(\mu)/C_2(\mu)\\
\end{array}\right.,
\label{eq:alphan-i}
\end{equation}
which follows from 
\begin{eqnarray}
\alpha_{N-i}&=\mu_i -i +\underbrace{2\kappa_{N-i}}_{\rm even}+\underbrace{N/2+1-\lambda}_{\rm even}\nonumber\\
&\equiv a'(i,\mu_i)+l'(i,\mu_i)\quad \mbox{mod }2. 
\end{eqnarray}
For $i\in [\mu'_1+1,N-1]$, we regard $(i,0)$ as an element of $C_2(\mu)$ ($D(\mu)\setminus C_2(\mu)$) when $i$ is even (odd). 
With use of (\ref{eq:alphan-i}), $E_{N-1}(\kappa)$ is rewritten as
\begin{equation}
E_{N-1}(\kappa)=\frac{\pi^2(1+2\lambda)^2}{L^2}\left({\cal E}'_{N-1}(\kappa)+{\cal E}''_{N-1}(\kappa)\right),
\label{eq:calE-def}
\end{equation}
with 
\begin{equation}
{\cal E}'_{N-1}(\kappa)=\sum_{i\in [1,N-1]; \mbox{\scriptsize \, s.t. } (i,\mu_i)\in C_2(\mu)}\left(i-\frac{N}{2}+\frac{\lambda-\mu_i +1}{2\lambda+1}\right)^2
\label{eq:Eprime1}
\end{equation}
and 
\begin{equation}
{\cal E}''_{N-1}(\kappa)=\sum_{i\in [1,N-1]; \mbox{\scriptsize \, s.t. } (i,\mu_i)\in D(\mu)\setminus C_2(\mu)}\left(i-\frac{N}{2}+\frac{\lambda-\mu_i }{2\lambda+1}\right)^2. 
\label{eq:Eprime2}
\end{equation}
Now we consider ${\cal E}'_{N-1}(\kappa)$. 
It is convenient to decompose the sum with respect to $i$ as
\begin{equation}
\sum_{i\in [1,N-1],\atop\mbox{\scriptsize \, s.t. } (i,\mu_i)\in C_2(\mu)}\rightarrow 
\sum_{j=0}^{2\lambda+1}\sum_{i\in [1+\mu'_{j+1},\mu'_j], \atop\mbox{\scriptsize \, s.t. } (i,\mu_i)\in C_2(\mu)}
\end{equation}
In the interval $i\in [1+\mu'_{j+1},\mu'_j]$, $i$'s satisfying $(i,\mu_i)\in C_2(\mu)$ align alternately. The minimum (maximum) value $i^{\square}_{\rm min}(j)$ ($i^{\square}_{\rm max}(j)$) of $i$ in $[1+\mu_{j+1},\mu_j]$ satisfying $(i,\mu_i)\in C_2(\mu)$  are given, respectively, by 
\begin{equation}
i^{\square}_{\rm min}(j)=\mu'_{j+1}+\sigma_{j+1}+3/2,\quad
i^{\square}_{\rm max}(j)=\mu'_{j}-\sigma_{j}-1/2,
\label{eq:imin-imax}
\end{equation}
as shown below. 
First we note that $(i,\mu_i)=(i,j)$ in the interval $i\in [1+\mu'_{j+1},\mu'_j]$. When $j+1\in Q$, $\mu'_{j+1}-(j+1)$ is even and hence $(1+\mu'_{j+1}, j)\in C_2(\mu)$ and $i^{\square}_{\rm min}(j)=1+\mu'_{j+1}$. When $j+1\in P$, on the other hand, $\mu'_{j+1}-(j+1)$ is odd and hence $(1+\mu'_{j+1}, j)\in D(\mu)\setminus C_2(\mu)$ and $i^{\square}_{\rm min}(j)=2+\mu'_{j+1}$. Thus we arrive at the first equation of (\ref{eq:imin-imax}). We can obtain the second equation of (\ref{eq:imin-imax}) in a similar way. In term of $i^{\square}_{\rm min}(j)$ and $i^{\square}_{\rm max}(j)$, ${\cal E}'_{N-1}(\kappa)$ is written as
\begin{eqnarray}
{\cal E}'_{N-1}(\kappa)
&=\sum_{j=0}^{2\lambda+1}\sum_{i=i^{\square}_{\rm min}(j), i^{\square}_{\rm min}(j)+2,\cdots}^{i^{\square}_{\rm max}(j)}
\left(i-\frac{N}{2}+\frac{\lambda-j+1}{2\lambda+1}\right)^2\nonumber\\
&=\frac16\sum_{j=0}^{2\lambda+1}\left(i^{\square}_{\rm max}(j)+1-N/2+\frac{\lambda-j+1}{2\lambda+1}\right)^3\nonumber\\
&\quad-\frac16\sum_{j=0}^{2\lambda+1}
 \left(i^{\square}_{\rm min}(j)-1-N/2+\frac{\lambda-j+1}{2\lambda+1}\right)^3\nonumber\\
&\quad+\frac16\sum_{j=0}^{2\lambda+1}\left(-i^{\square}_{\rm max}(j)+i^{\square}_{\rm min}(j)-2\right). 
\label{eq:Eprime1-re}
\end{eqnarray}
In the second equality, we have used following formula:
\begin{eqnarray}
&\sum_{i=m,m+2,m+4,\cdots, n} (i+A)^2\nonumber\\
&=\frac16\left[(n+1+A)^3-(m-1+A)^3-n+m-2\right]
\label{eq:square-sum-formula}
\end{eqnarray}
for $n$, $m$ satisfying $n-m$ being a positive and even integer. 
The expression (\ref{eq:Eprime1-re}) can be further rewritten as
\begin{eqnarray}
{\cal E}'_{N-1}(\kappa)
&=\frac16\sum_{j=1}^{2\lambda+1}\left[\left(\tilde{\mu}'_j-\sigma_j\right)^3-\left(\tilde{\mu}'_j+\sigma_j+1/(2\lambda+1)\right)^3+2\sigma_j\right]\nonumber\\
&+\frac{M(M^2-1)}{3}+\frac{\lambda^2 M}{(2\lambda+1)^2}+\frac16 
\label{eq:Eprime1-final}
\end{eqnarray}
with use of (\ref{eq:imin-imax}) and (\ref{eq:mutildej}). 
We can rewrite the expression (\ref{eq:Eprime2}) for ${\cal E}''_{N-1}(\kappa)$ in a similar way. First (\ref{eq:Eprime2}) is rewritten as
\begin{eqnarray}
{\cal E}''_{N-1}(\kappa)&=\sum_{j=0}^{2\lambda+1}\sum_{i\in [1+\mu'_{j+1},\mu'_j], \atop\mbox{\scriptsize \, s.t. } (i,\mu_i)\in D(\mu)\setminus C_2(\mu)}\left(i-\frac{N}{2}+\frac{\lambda-j}{2\lambda+1}\right)^2\nonumber\\
&=\sum_{j=0}^{2\lambda+1}\sum_{i=i^{\blacksquare}_{\rm min}(j), i^{\blacksquare}_{\rm min}(j)+2,\cdots}^{i^{\blacksquare}_{\rm max}(j)}
\left(i-\frac{N}{2}+\frac{\lambda-j}{2\lambda+1}\right)^2,\nonumber\\
&=\frac16\sum_{j=0}^{2\lambda+1} \left(i^{\blacksquare}_{\rm max}(j)+1-N/2+\frac{\lambda-j}{2\lambda+1}\right)^3\nonumber\\
&\quad -\frac16\sum_{j=0}^{2\lambda+1} 
 \left(i^{\blacksquare}_{\rm min}(j)-1-N/2+\frac{\lambda-j}{2\lambda+1}\right)^3\nonumber\\
&\quad+\frac16\sum_{j=0}^{2\lambda+1}\left(-i^{\blacksquare}_{\rm max}(j)+i^{\blacksquare}_{\rm min}(j)-2\right)
\label{eq:Eprime2-re}
\end{eqnarray}
where 
\begin{equation}
i^{\blacksquare}_{\rm min}(j)=\mu'_{j+1}-\sigma_{j+1}+3/2,\quad i^{\blacksquare}_{\rm max}(j)=\mu'_j+\sigma_j-1/2
\label{eq:i-black-min-max}
\end{equation}
 are, respectively, defined as the minimum and maximum of $i$ satisfying $(i,\mu_i)\in D(\mu)/C_2(\mu)$ in $[1+\mu'_{j+1},\mu'_j]$. In the last equality in (\ref{eq:Eprime2-re}), we have used (\ref{eq:square-sum-formula}). 
Substituting (\ref{eq:i-black-min-max}) into (\ref{eq:Eprime2-re}) and using (\ref{eq:mutildej}), ${\cal E}''_{N-1}(\kappa)$ is expressed as
\begin{eqnarray}
{\cal E}''_{N-1}(\kappa)
&=\frac16\sum_{j=1}^{2\lambda+1}\left[\left(\tilde{\mu}'_j+\sigma_j-1/(2\lambda+1)\right)^3-\left(\tilde{\mu}'_j-\sigma_j\right)^3-2\sigma_j\right]\nonumber\\
&+\frac{M(M^2-1)}{3}+\frac{\lambda^2 M}{(2\lambda+1)^2}+\frac16. 
\label{eq:Eprime2-final}
\end{eqnarray}
Substituting (\ref{eq:Eprime1-final}) and (\ref{eq:Eprime2-final}) into (\ref{eq:calE-def}), we obtain 
\begin{eqnarray}
E_{N-1}(\kappa)&=-(2\lambda+1)\sum_{j=1}^{2\lambda+1}\left(\frac{\pi(\tilde{\mu}'_j+\sigma_j)}{L}\right)^2+
\frac{4\pi^2\lambda(\lambda+1)}{3L^2}\nonumber\\
&\quad+\underbrace{\frac{2\pi^2(2\lambda+1)^2M(M^2-1)}{3L^2}+\frac{2\pi^2\lambda^2 M}{L^2}}_{=E_{N}({\rm g})}\end{eqnarray}
from which and (\ref{eq:omegakappa}), the expressions (\ref{eq:tildeomega}) and (\ref{eq:tildeP}) follow. 

\subsection{Matrix elements}
\label{Me-ME}
Here we describe $X_{\mu},Y_{\mu}(r),Z_{\mu}(r)$ in the matrix elements
in terms of the renormalized momenta (\ref{eq:mutildej})
and the spin variables (\ref{eq:sigma-j}) and (\ref{eq:sigma0-2lambda+1}).
First we consider $X_\mu$. 
The product with respect to $s=(i,j)\in D(\mu)\setminus C_2(\mu)$ 
is taken within each column, and then taken over the column. 
In the $j$-th column, this condition
is equivalent to even (odd) $i$ when $j$ is odd (even).
The square $s=(\mu'_j,j)$ belongs to $D(\mu)\setminus C_2(\mu)$ 
when $j\in P$ and $s=(\mu'_j -1,j)$ belongs to $D(\mu)\setminus C_2(\mu)$ 
when $j\in Q$. 
The maximum value of $i$ in the $j$-th column is thus expressed as
\begin{equation}
i_{\rm max}=\mu'_j -\delta_{\sigma_j\sigma_{2\lambda+2}}. 
\label{eq:imax-muj}
\end{equation}
%
The contribution to $X_\mu$ from $j$-th column is then given by
\begin{eqnarray}
\prod_{i: {\rm even}}^{i_{\rm max}}
\left(-\alpha(j-1)+i/2\right)
&=\prod_{i'=1}^{i_{\rm max}/2}(i'-\alpha(j-1))\nonumber\\
&=\frac{\Gamma[i_{\rm max}/2+1-\alpha(j-1)]}{\Gamma[1-\alpha(j-1)]}
\label{eq:Xjodd}
\end{eqnarray}
for odd $j$ and
\begin{eqnarray}
\prod_{i: {\rm odd}}^{i_{\rm max}}
\left(-\alpha(j-1)+i/2\right)
&=\prod_{i'=1}^{(i_{\rm max}+1)/2}(i'-1/2-\alpha(j-1))\nonumber\\
&=\frac{\Gamma[i_{\rm max}/2+1-\alpha(j-1)]}{\Gamma[1/2-\alpha(j-1)]}
\label{eq:Xjeven}
\end{eqnarray}
for even $j$. We introduce a dummy index $i'=i/2$ in (\ref{eq:Xjodd})
 and $i'=(i+1)/2$ in (\ref{eq:Xjeven}), respectively. 
 From (\ref{eq:Xjodd}) and (\ref{eq:Xjeven}) and with use of 
\begin{eqnarray}
&\prod_{j\in [1,3,\cdots,2\lambda+1]}\Gamma[1-\alpha(j-1)]
\prod_{j\in [2,4,\cdots,2\lambda]}\Gamma[1/2-\alpha(j-1)]\nonumber\\
&=\prod_{j=1}^{2\lambda+1}\Gamma[j/(2\lambda+1)],
\end{eqnarray}
we obtain 
\begin{equation}
X_\mu=\prod_{j=1}^{2\lambda+1}
\frac{\Gamma[i_{\rm max}/2+1-\alpha(j-1)]}{\Gamma[j/(2\lambda+1)]}.
\label{eq:Xmu-imd}
\end{equation}
With use of (\ref{eq:imax-muj}) and 
(\ref{eq:mutildej}), 
(\ref{eq:Xmu-imd}) becomes (\ref{eq:Xmu-final}).

Similarly, we can write $Y_\mu(r)$ in terms of $\tilde{\mu}'_j$ 
and $\sigma_j$ for $j\in [0,2\lambda+1]$, as shown below. 
When $(i,j)\in C_2(\mu)$, both $i$ and $j$ are odd or even. 
The maximum value $i_{\rm max}$ of $i$ 
in $j$-th column is given by $i_{\rm max}=\mu'_j-\delta_{\sigma_j\sigma_0}$. 
The contribution to $Y_\mu(r)$ from $j$-th column is given by
\begin{eqnarray}
\prod_{i: {\rm odd}}^{i_{\rm max}}(\alpha(j-1)+r+(N-i)/2)
&=\prod_{i'=1}^{(i_{\rm max}+1)/2}(\alpha(j-1)+r+(N+1)/2-i')\nonumber\\
&=\frac{\Gamma[\alpha(j-1)+r+(N+1)/2]}{\Gamma[\alpha(j-1)+r+(N-i_{\rm max})/2]}
\label{eq:Ymujodd}
\end{eqnarray}
for odd $j$ and
\begin{eqnarray}
\prod_{i: {\rm even}}^{i_{\rm max}}(\alpha(j-1)+r+(N-i)/2)
&=\prod_{i'=1}^{i_{\rm max}/2}(\alpha(j-1)+r+N/2-i')\nonumber\\
&=\frac{\Gamma[\alpha(j-1)+r+N/2]}{\Gamma[\alpha(j-1)+r+(N-i_{\rm max})/2]}
\label{eq:Ymujeven}
\end{eqnarray}
for even $j$. 
With use of (\ref{eq:mutildej}) and 
the relation
\begin{eqnarray}
&\prod_{j\in [1,3,\cdots,2\lambda+1]}\Gamma[\alpha(j-1)+r+(N+1)/2]
\prod_{j\in [2,4,\cdots,2\lambda]}\Gamma[\alpha(j-1)+r+N/2]\nonumber\\
&=\prod_{j=1}^{2\lambda+1}\Gamma[\alpha+r+N/2+(j-1)/(2\lambda+1)],
\end{eqnarray}
the expression $Y_\mu(r)$ is obtained as (\ref{eq:Ymu-final}).

\begin{figure}
\begin{center}
\input{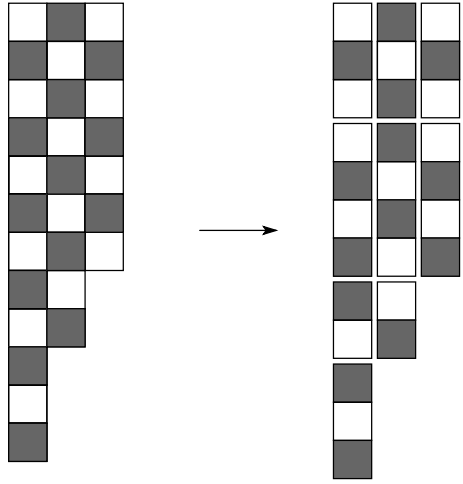}
\caption{Young diagram of a quasi-hole state for $\lambda=1$ is decomposed into $D_{jk}$ for calculation of $Z_\mu(r)$. }
\label{fig:qhyoungdecomposition}
\end{center}
\end{figure}

Next we consider the expression for $Z_\mu(r)$ in terms of 
$\{\tilde{\mu}'_j, \sigma_j\}$. 
It is convenient to decompose $D(\mu)$ as shown 
in Figure~\ref{fig:qhyoungdecomposition}. 
Correspondingly, $Z_\mu(r)$ is rewritten as
\begin{equation}
Z_\mu(r)=\prod_{j=1}^{2\lambda+1}\prod_{k=j}^{2\lambda+1}
\prod_{s\in H_2(\mu)\cap D_{jk}}\left(\alpha a(s)+r+l(s)/2\right)
\label{eq:a-l-decomposition}
\end{equation}
with 
\begin{equation}
D_{jk}=\left\{s=(i,j)|i\in [1+\mu'_{k+1}, \mu'_k]\right\}.  
\end{equation}
Within $D_{jk}$, the arm length $a(s)$ is constant ($=k-j$) 
and hence the squares which belong to $H_2(\mu)$ 
and $D(\mu)\setminus H_2(\mu)$ are, respectively, aligned alternately. 
Let $i_{\rm min}$ and $i_{\rm max}$ be, respectively, 
the minimum and maximum values of $i$ in $s\in H_2(\mu)\cap D_{jk}$. 
Thus the product over $s=(i,j)\in H_2(\mu)\cap D_{jk}$ 
is expressed as the product over 
$i'=0,1,\cdots, (i_{\rm max}-i_{\rm min})/2$ with $i=2i'+i_{\rm min}$. 
The contribution to $Z_\mu(r)$ from $D_{jk}$ is given by
\begin{eqnarray}
&\prod_{s\in H_2(\mu)\cap D_{jk}}\left(\alpha a(s)+r+l(s)/2\right)\nonumber\\
&=\sum_{i'=0}^{(i_{\rm max}-i_{\rm min})/2}
(\alpha(k-j)+r+\mu'_j/2-i_{\rm min}/2-i')\nonumber\\
&=\frac{\Gamma[\alpha(k-j)+r+\mu'_j/2-i_{\rm min}/2+1]}
{\Gamma[\alpha(k-j)+r+\mu'_j/2-i_{\rm max}/2]}.
\label{eq:Zmu-Djk-imd}
\end{eqnarray}
The maximum value $i_{\rm max}$ of $i$ in $H_2(\mu)\cap D_{jk}$ is expressed as
\begin{equation}
i_{\rm max}=\mu'_k-\delta_{\sigma_j\sigma_k}
\label{eq:i-max}
\end{equation}
as shown below. 

For $s\in D_{jk}$, $a(s)$ is $k-j$ 
and $l(s)=\mu'_j-i$ and hence $h(s)=a(s)+l(s)+1$ is written as
\begin{eqnarray}
h(s)=\mu'_j-j+k-i+1=\mu'_j-j-(\mu'_k-k)+\mu'_k-i+1.
\label{eq:hofs}
\end{eqnarray}
When $(j,k)\in (P,P)$ or $(Q,Q)$, both $\mu'_j-j$ 
and $\mu'_k-k$ are odd or even and hence for $(i,j)\in H_2(\mu)\cap D_{jk}$,
\begin{equation}
\mu'_k-1\equiv i\quad {\rm mod}\, 2.
\label{eq:imax-jkPP}
\end{equation}
When $(j,k)\in (P,Q)$ or $(Q,P)$, on the other hand, 
the sum of $\mu'_j-j$ and $\mu'_k-k$ is odd and hence,
\begin{equation}
\mu'_k\equiv i\quad {\rm mod}\, 2.
\label{eq:imax-jkPQ}
\end{equation}
The two relations (\ref{eq:imax-jkPP}) and (\ref{eq:imax-jkPQ}) 
are expressed as (\ref{eq:i-max}) in a unified way. 
In a similar way, $i_{\rm min}$ is expressed as 
\begin{equation}
i_{\rm min}=\mu'_{k+1}+1+\delta_{\sigma_j\sigma_{k+1}}
\label{eq:i-min}
\end{equation}
respectively. 

The expression (\ref{eq:Zmu-Djk-imd}) is further written as
\begin{eqnarray}
&\prod_{s\in H_2(\mu)\cap D_{jk}}\left(\alpha a(s)+r+l(s)/2\right)\nonumber\\
&=\frac{\Gamma[(\tilde{\mu}'_j -\tilde{\mu}'_{k+1}-
\delta_{\sigma_j\sigma_{k+1}})/2 +r-\alpha+1/2]}
{\Gamma[(\tilde{\mu}'_j -\tilde{\mu}'_{k}+\delta_{\sigma_j\sigma_{k}})/2 +r]},
\label{eq:Zmu-Djk}
\end{eqnarray}
with use of (\ref{eq:i-max}), (\ref{eq:i-min}) 
and  (\ref{eq:mutildej}).  
From (\ref{eq:Zmu-Djk}), the expression (\ref{eq:zmu-r}) for $Z_\mu(r)$  follows. 
\section{Spectral Weight}\label{SW}

\begin{figure}
 \begin{center}
  \begin{tabular}{cc}
   \begin{minipage}{0.5\hsize}
    \includegraphics[width=80mm]{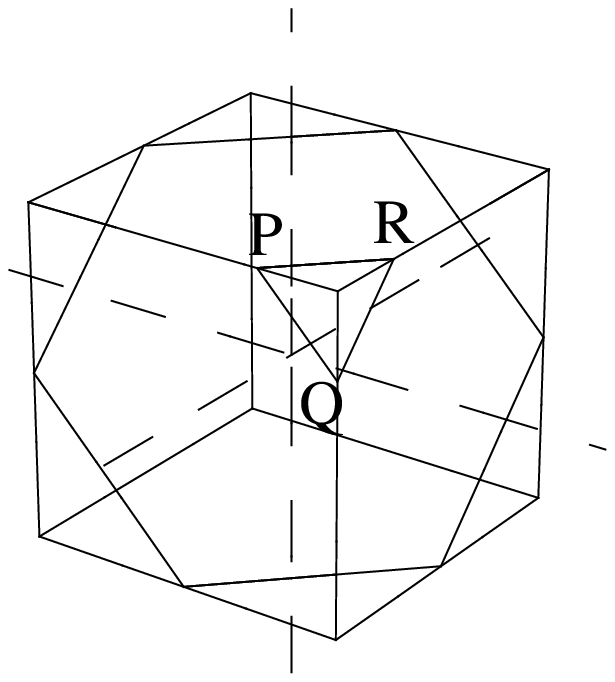}
    \vspace{-14mm}
 \caption{Cross section of the plane (\ref{eq:momentum-c}) and the cube   (\ref{cube-c}). The cross section is triangular PQR when $p\in [\pi d/2,3\pi d/2]$ and hexagon when $p\in [0,\pi d/2]$. }
 \label{fig:cube}
   \end{minipage}
   \begin{minipage}{0.5\hsize}
    \includegraphics[width=50mm]{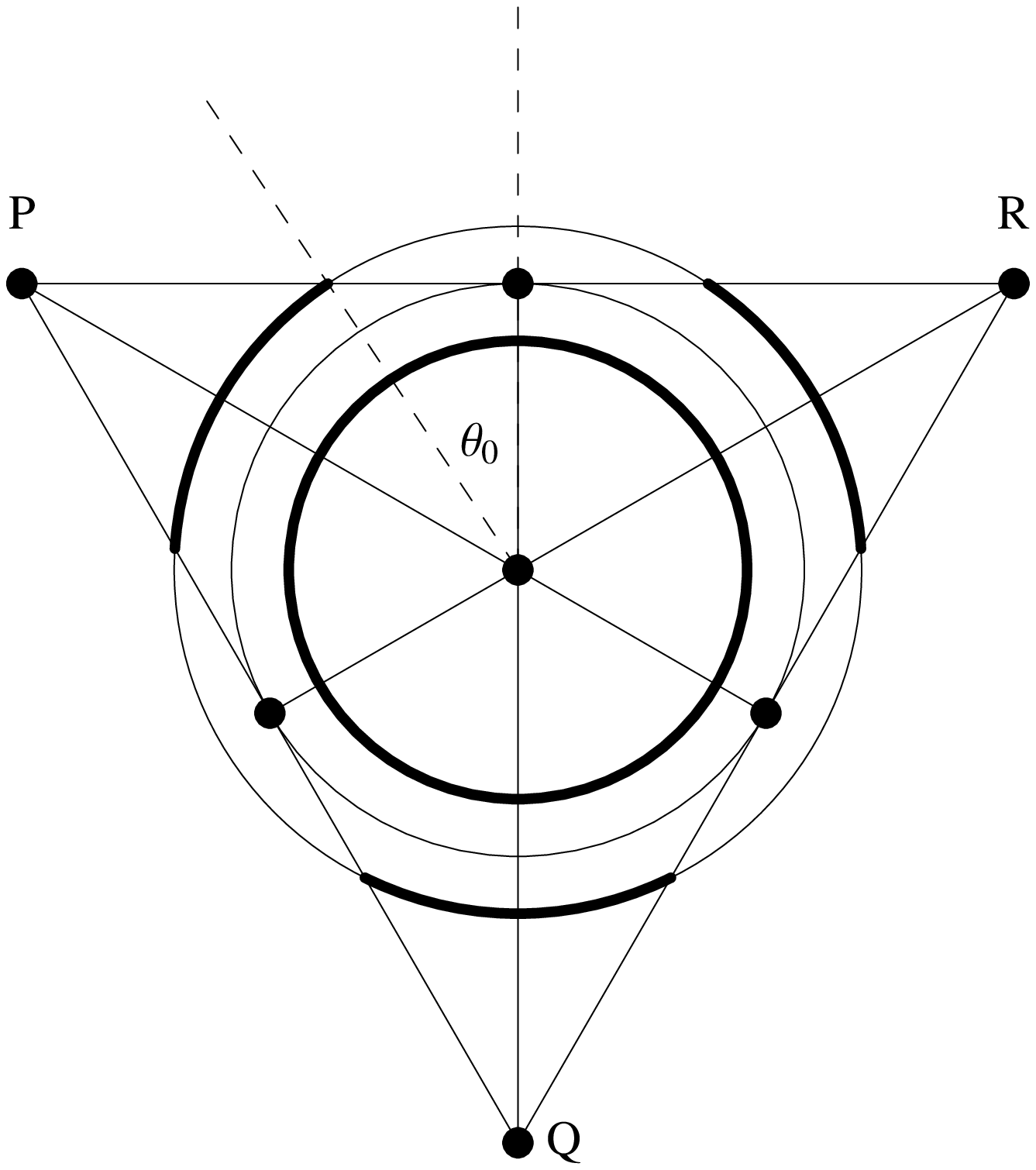}
 \caption{Cross section of the plane (\ref{eq:momentum-c}) and the cube   (\ref{cube-c}) for $p\in [\pi d/2,3\pi d/2]$. }
 \label{fig:triangular}
   \end{minipage}
  \end{tabular}
 \end{center}
\end{figure}

The triple integral in the spectral function for $\lambda=1$,
\begin{eqnarray}
 A^{-}(\epsilon,p)
 &=C
 \int_{-1}^{1}\rmd u_1 \int_{-1}^{1}\rmd u_2 \int_{-1}^{1}\rmd u_3
 \,\,\delta\left(p-\frac{\pi d}{2}(u_1+u_2+u_3)\right)
 \nonumber\\
 &\,\,\,\quad\times
 \delta\!\left(\epsilon-\frac{3(\pi d)^2}{4}\left(3-u_1^2-u_2^2-u_3^2\right)
 \right) F(u_1,u_2,u_3)
 \label{sf}
\end{eqnarray}
with  
\begin{eqnarray}
&F(u_1,u_2,u_3)\nonumber\\
&=
 \left|u_2-u_3\right|^{4/3}
 \left|u_1-u_3\right|^{-2/3}\left|u_1-u_2\right|^{-2/3} \prod_{j=1}^{3}
 \left(1-u_j^2\right)^{-1/3}
\end{eqnarray}
reduces to an integral on a curve determined by the sphere, plane,
and the cube
\begin{eqnarray}
&\epsilon=\frac{3(\pi d)^2}{4}\left(3-u_1^2-u_2^2-u_3^2\right)
\label{eq:energy-c} \\
&p=\frac{\pi d}{2}(u_1+u_2+u_3)
\label{eq:momentum-c} \\
&|u_i|\le 1,\quad i=1,2,3.
\label{cube-c}
\end{eqnarray}
The cross section between (\ref{eq:momentum-c}) and (\ref{cube-c}) is triangular when
\begin{equation}
\pi d/2\le p\le 3\pi d/2
\label{eq:triangular}
\end{equation}
and hexagon when 
\begin{equation}
0\le p\le \pi d/2, 
\label{eq:hexagon}
\end{equation}
as shown in Figure~\ref{fig:cube}.

In the following, we consider the case (\ref{eq:triangular}) only. The case (\ref{eq:hexagon}) can be discussed similarly. 
The integral in (\ref{sf}) is on a circle
when  
\begin{equation}
\frac{\pi d}{2}\le p\le \frac{3\pi d}{2},\mbox{ and } 
-\frac{3}{2}\left(p-\frac{\pi d}{2}\right)^2+\frac32(\pi d)^2\le \epsilon 
\le -p^2+\frac94 (\pi d)^2,
\end{equation}
and it is on three disconnected pieces of arc when 
$$
\pi d/2\le p\le 3\pi d/2
$$
and
$$-3(p-\pi d)^2+\frac{3}{4}(\pi d)^2\le \epsilon 
\le -\frac{3}{2}\left(p-\frac{\pi d}{2}\right)^2+\frac32(\pi d)^2,
$$
as shown by bold curves in Figure~\ref{fig:triangular}.  
The integrand $F(u_1,u_2,u_3)$ in (\ref{sf}) diverges when 
\begin{equation}
u_1=u_3,\quad u_1=u_2\quad \mbox{or } |u_j|=1\mbox{ for }j=1,2,3, 
\label{eq:singular-lines}
\end{equation}
which are represented by solid lines in Figure~\ref{fig:triangular}. When the contour of the integral passes near the crossing points of the above lines (\ref{eq:singular-lines}), the spectral function becomes singular. Those crossing points are given by
\begin{eqnarray}
 (u_1,u_2,u_3)=
%
 &\left(\frac{2p}{3\pi d }\,,\,
 \frac{2p}{3\pi d }\,,\,\frac{2p}{3\pi d }\right)\label{eq:crosspoint-upper-edge},\\
 &\left(1\,,\,
        1\,,\,
        -2+\frac{2p}{\pi d }\right),\label{eq:crosspoint-lower-edge}\\
 &\left(-\frac{1}{2}+\frac{p}{\pi d}\,,\,
 -\frac{1}{2}+\frac{p}{\pi d}\,,\,1\right)\label{eq:crosspoint-middle}
\end{eqnarray}
and their equivalent points. They are depicted in Figure~\ref{fig:triangular} by dots. The singularity of the spectral function near the upper edge $\epsilon=(3\pi d/2)^2(1-(2p/(\pi d))^2)$ of the support comes from (\ref{eq:crosspoint-upper-edge}). The point (\ref{eq:crosspoint-lower-edge}) yields the singularity of $A(\epsilon,p)$ near the lower edge $\epsilon=3(\pi d/2)^2(2p/(\pi d)-1)(3-2p/(\pi d)))$. (\ref{eq:crosspoint-middle}) is relevant to the singularity near 
\begin{equation}
\epsilon=
\frac32\left(p+\frac{\pi d}{2}\right)\left(\frac{3\pi d}{2}-p\right)
\equiv \epsilon_p.\label{eq:two-qh}
\end{equation}
The singularities near the upper edge and lower edge have been obtained in earlier papers\cite{KYA1997,Yamamoto1999}. We thus consider the singularity when $\delta\epsilon\equiv \epsilon-\epsilon_p\sim 0$ in the following. 

\begin{figure}
\begin{center}
\unitlength 0.1in
\begin{picture}( 17.9000, 17.9000)(  5.9300,-23.9000)
%
\special{pn 8}%
\special{pa 1488 1496}%
\special{pa 1488 600}%
\special{fp}%
\special{sh 1}%
\special{pa 1488 600}%
\special{pa 1468 668}%
\special{pa 1488 654}%
\special{pa 1508 668}%
\special{pa 1488 600}%
\special{fp}%
%
\special{pn 8}%
\special{pa 1488 1496}%
\special{pa 594 1496}%
\special{fp}%
\special{sh 1}%
\special{pa 594 1496}%
\special{pa 660 1516}%
\special{pa 646 1496}%
\special{pa 660 1476}%
\special{pa 594 1496}%
\special{fp}%
%
\special{pn 8}%
\special{sh 0}%
\special{ar 1488 1496 84 84  0.0000000 6.2831853}%
\put(16.0300,-15.0800){\makebox(0,0)[lb]{$\bmath{e}_\zeta$}}%
\put(16.0300,-9.9600){\makebox(0,0)[lb]{$\bmath{e}_\xi$}}%
\put(9.6000,-14.7000){\makebox(0,0)[lb]{$\bmath{e}_\eta$}}%
%
\special{pn 8}%
\special{ar 1488 1496 896 896  0.0000000 6.2831853}%
%
\special{pn 4}%
\special{pa 1488 1496}%
\special{pa 1008 742}%
\special{fp}%
\special{sh 1}%
\special{pa 1008 742}%
\special{pa 1028 808}%
\special{pa 1038 786}%
\special{pa 1062 786}%
\special{pa 1008 742}%
\special{fp}%
%
\special{pn 4}%
\special{pa 1488 1496}%
\special{pa 734 1974}%
\special{fp}%
\special{sh 1}%
\special{pa 734 1974}%
\special{pa 800 1956}%
\special{pa 778 1946}%
\special{pa 780 1922}%
\special{pa 734 1974}%
\special{fp}%
%
\special{pn 8}%
\special{sh 0}%
\special{ar 1488 1496 90 90  0.0000000 6.2831853}%
%
\special{pn 4}%
\special{sh 0.300}%
\special{ar 1488 1496 46 46  0.0000000 6.2831853}%
\put(13.7000,-12.9000){\makebox(0,0)[lb]{$\theta$}}%
\put(9.6400,-11.2400){\makebox(0,0)[lb]{$\bmath{e}_\rho$}}%
\put(10.6000,-18.9000){\makebox(0,0)[lb]{$\bmath{e}_\theta$}}%
\put(11.4000,-16.7000){\makebox(0,0)[lb]{$\theta$}}%
\end{picture}%
\end{center}
\caption{Cartesian coordinate $\{\bmath{e}_\xi,\bmath{e}_\eta,\bmath{e}_\zeta\}$ and cylindrical coordinate $\{\bmath{e}_\rho,\bmath{e}_\theta,\bmath{e}_\zeta\}$.} 
\label{fig:cylinder}
\end{figure}
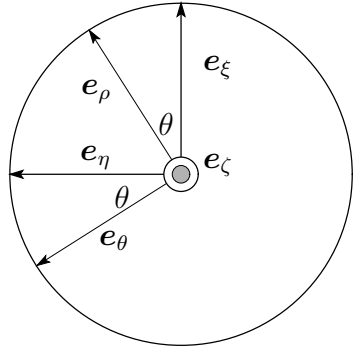

Now we introduce the Cartesian coordinate $\bmath{e}_i\cdot \bmath{e}_j=\delta_{ij}$ $i,j=1,2,3$ and define the vector
\begin{equation}
\bmath{u}=u_1\bmath{e}_1+u_2\bmath{e}_2+u_3\bmath{e}_3.
\label{eq:u-vector}
\end{equation}
We define another Cartesian coordinate
\begin{eqnarray}
&\bmath{e}_\xi=-\bmath{e}_1/\sqrt{6}-\bmath{e}_2/\sqrt{6}+2\bmath{e}_3/\sqrt{6},\nonumber\\
&\bmath{e}_\eta=-\bmath{e}_1/\sqrt{2}+\bmath{e}_2/\sqrt{2},\nonumber\\
&\bmath{e}_\zeta=(\bmath{e}_1+\bmath{e}_2+\bmath{e}_3)/\sqrt{3}
\label{eq:another-cart}
\end{eqnarray}
and circular coordinate on $\bmath{e}_\xi-\bmath{e}_\eta$ plane
\begin{eqnarray}
&\bmath{e}_\rho=\bmath{e}_\xi \cos\theta 
+\bmath{e}_\eta \sin \theta,\nonumber\\
&\bmath{e}_\theta=-\bmath{e}_\xi\sin \theta 
+\bmath{e}_\eta \cos\theta. \label{eq:circular-c}
\end{eqnarray}
Note that $\{\bmath{e}_\xi, \bmath{e}_\eta,\bmath{e}_\zeta\}$ forms a cylindrical coordinate as shown in Figure~\ref{fig:cylinder}.
In terms of (\ref{eq:another-cart}) and (\ref{eq:circular-c}), we rewrite $\bmath{u}$ as
\begin{equation}
\bmath{u}=u_\rho \bmath{e}_\rho +u_\zeta\bmath{e}_\zeta.
\label{eq:u-rho-zeta}
\end{equation}
In the following, we change the variables of integral in (\ref{sf}) from $(u_1,u_2,u_3)$ to $(u_\rho,\theta,u_\zeta)$. 
From (\ref{eq:u-vector})-(\ref{eq:u-rho-zeta}), we obtain 
\begin{eqnarray}
&u_1=\bmath{u}\cdot\bmath{e}_1=u_\rho (-\cos\theta /\sqrt{6}-\sin\theta /\sqrt{2})+u_\zeta/\sqrt{3}\nonumber\\
&u_2=\bmath{u}\cdot\bmath{e}_2=u_\rho (-\cos\theta /\sqrt{6}+\sin\theta /\sqrt{2})+u_\zeta/\sqrt{3}\nonumber\\
&u_3=\bmath{u}\cdot\bmath{e}_3=2 u_\rho \cos\theta /\sqrt{6}+u_\zeta/\sqrt{3}.
\end{eqnarray}
The spectral function (\ref{sf}) is rewritten as
\begin{eqnarray}
 A^{-}(\epsilon,p)
 &=C
 \int \rmd u_\rho u_\rho \int \rmd\theta \int \rmd u_\zeta
 \,\,\delta\left(p-\frac{\sqrt{3}\pi d u_\zeta}{2}\right)
 \nonumber\\
 &\,\,\,\quad\times
 \delta\!\left(\epsilon-3(\pi d/2)^2\left(3-u_\rho^2-u_\zeta^2\right)
 \right)
 \nonumber\\
 &\,\,\,\quad\times
F(u_1,u_2,u_3).
 \label{sf-rewritten}
\end{eqnarray}
From the delta functions, $u_\rho$ and $u_\zeta$ are forced to be
\begin{equation}
u_\rho= \sqrt{3-4(\epsilon+p^2)/(3\pi^2 d^2)},\quad
u_\zeta=2p/(\sqrt{3}\pi d), 
\end{equation}
respectively. As a result,  $A^{-}(\epsilon,p)$ becomes
\begin{equation}
A^{-}(\epsilon,p)=\frac{4C}{(\sqrt{3}\pi d)^3}\int\rmd \theta F(\bar{u}_1(\theta),\bar{u}_2(\theta),\bar{u}_3(\theta))
\label{eq:A-theta}
\end{equation}
with 
\begin{eqnarray}
&\bar{u}_1(\theta)= \sqrt{(1/2-p'/3)^2-2\delta\epsilon'/9}(-\cos\theta -\sqrt{3}\sin\theta )+2p'/3\nonumber\\
&\bar{u}_2(\theta)= \sqrt{(1/2-p'/3)^2-2\delta\epsilon'/9}(-\cos\theta+\sqrt{3}\sin\theta )+2p'/3\nonumber\\
&\bar{u}_3(\theta)= \sqrt{(1-2p'/3)^2-4\delta\epsilon'/9}\cos\theta +2p'/3.
\end{eqnarray}
Here we have introduced $\delta\epsilon'\equiv \delta\epsilon/(\pi d)^2$ 
and $p'=p/(\pi d)$. 
The integral in (\ref{eq:A-theta}) runs over
\begin{eqnarray}
&\theta\in [0,2\pi], \quad\mbox{for}\quad\delta\epsilon >0\nonumber\\
&\theta\in [\theta_0,2\pi/3-\theta_0]\cup[2\pi/3+\theta_0,4\pi/3-\theta_0]\cup[4\pi/3+\theta_0,2\pi-\theta_0] 
\end{eqnarray}
for $\delta\epsilon <0$ with $$
\theta_0=\arcsin
\left[\sqrt{\frac{-2\delta\epsilon'}{(3/2-p')^2-2\delta\epsilon'}}\,\right]. 
$$ 
The point (\ref{eq:crosspoint-middle}) and equivalent points
\begin{equation}
\left(1\,,\,
 -\frac{1}{2}+\frac{p}{\pi d}\,,\,-\frac{1}{2}+\frac{p}{\pi d}\right),\quad
\left(-\frac{1}{2}+\frac{p}{\pi d}\,,\,
 1\,,\,-\frac{1}{2}+\frac{p}{\pi d}\right)
\end{equation}
correspond to $\delta\epsilon'=0$ and $\theta=0,\, 2\pi/3,\, 4\pi/3$, respectively. The most singular contribution to $A^-(\epsilon,p)$ for $\delta\epsilon=0$ comes from the vicinity of $\theta=0$, where the factors $|u_1 -u_2|^{-2/3}(1-u_3)^{-1/3}$ become singular in $F(u_1,u_2,u_3)$. Therefore, we approximate $F(u_1,u_2,u_3)$ as
\begin{eqnarray}
F&\propto (\bar{u}_1(\theta)-\bar{u}_2(\theta))^{-2/3}(1-\bar{u}_3(\theta))^{-1/3}\propto \frac{|\sin \theta|^{-2/3}}{(1-g \cos\theta)^{1/3}}
\end{eqnarray}
with $g=\sqrt{1-\delta\epsilon'/(p'-3/2)^2}$. $A^-(\epsilon,p)$ is evaluated as 
 \begin{equation}
 A^-(\epsilon,p)\propto \int_0^{\theta_{\rm c}}\frac{|\sin\theta|^{-2/3}}{(1-g\cos\theta)^{1/3}}\rmd \theta
\label{eq:a-singular+}
 \end{equation}
 for $\delta\epsilon>0$ and $\delta\epsilon\sim 0$
 and 
 \begin{equation}
 A^-(\epsilon,p)\propto \int_{\theta_0}^{\theta_{\rm c}}\frac{|\sin\theta|^{-2/3}}{(1-g\cos\theta)^{1/3}}\rmd \theta
\label{eq:a-singular-}
 \end{equation}
 for $\delta\epsilon<0$ and $\delta\epsilon\sim 0$.
Here $\theta_{\rm c}$ is a cut-off angle of the order unity. We can take $\theta_{\rm c}=\pi/3,$ or $\pi/2$, for example. 
First we evaluate (\ref{eq:a-singular+}). 
Introducing $t=\tan(\theta/2)$, $t_{\rm c}=\tan(\theta_{\rm c}/2)$ and $\tilde{g}=(1-g)/(1+g)$,  (\ref{eq:a-singular+}) becomes 
\begin{equation}
 A^-(\epsilon,p)\propto \int_0^{t_{\rm c}}\frac{t^{-2/3}(1+t^2)^{-4/3}}{(t^2 +\tilde{g})^{1/3}}\rmd t.
\label{eq:a-singular-2}
\end{equation}
We rewrite (\ref{eq:a-singular-2}) as
\begin{eqnarray}
&\int_0^{t_{\rm c}}\frac{t^{-2/3}(1+t^2)^{-4/3}}{(t^2 +\tilde{g})^{1/3}}\rmd t\\
&=\int_0^{t_{\rm c}}\frac{t^{-2/3}}{(t^2 +\tilde{g})^{1/3}}\rmd t+
\int_0^{t_{\rm c}}\frac{t^{-2/3}((1+t^2)^{-4/3}-1)}{(t^2 +\tilde{g})^{1/3}}\rmd t
\end{eqnarray}
The second integral of the right-hand side converges when $\tilde{g}=0$. The first term of the right-hand side, on the other hand, is rewritten as
\begin{equation}
\tilde{g}^{-1/6}\int_0^{t_{\rm c}\tilde{g}^{-1/2}}\frac{\tau^{-2/3}}{(\tau^2+1)^{1/3}}\rmd \tau
\sim \tilde{g}^{-1/6}\int_0^\infty\frac{\tau^{-2/3}}{(\tau^2+1)^{1/3}}\rmd \tau.
\end{equation}
Since $\tilde{g}\propto \delta\epsilon$, we arrive at
\begin{equation}
A^-(\epsilon,p)\propto (\delta \epsilon)^{-1/6}
\end{equation}
when $\delta \epsilon>0$ and $\delta \epsilon\sim 0$.

The singularity near $\delta\epsilon=0$ and $\delta\epsilon<0$ is evaluated similarly. 
Introducing $t_0=\tan(\theta_0/2)$, (\ref{eq:a-singular-}) becomes
\begin{eqnarray}
 A^-(\epsilon,p)&\propto \int_{t_0}^{t_{\rm c}}\frac{t^{-2/3}(1+t^2)^{-4/3}}{(t^2 +\tilde{g})^{1/3}}\rmd t\sim \tilde{g}^{-1/6}\int_{t_0\tilde{g}^{-1/2}}^{t_{\rm c}\tilde{g}^{-1/2}}\frac{\tau^{-2/3}}{(\tau^2 +1)^{1/3}}\rmd \tau
\label{eq:a-singular-2-}
\end{eqnarray}
When $\tilde{g}\sim0$, equivalently $\delta\epsilon\sim 0$, $t_0\tilde{g}^{-1/2}$ is the order of unity and hence,
\begin{equation}
 A^-(\epsilon,p)\propto (-\delta\epsilon)^{-1/6}
\end{equation}
when $\delta \epsilon<0$ and $\delta \epsilon\sim 0$.

\end{document}